\documentclass[a4paper,fleqn]{cas-sc}

\usepackage[numbers, sort&compress]{natbib}
\usepackage{amssymb}
\usepackage{amsmath}
\usepackage{hyperref}
\usepackage{caption}
\usepackage{subcaption}
\usepackage{xcolor}
\usepackage{lipsum}
\usepackage[normalem]{ulem} % Para tachar texto
\usepackage{xcolor}         % Para el color

\def\tsc#1{\csdef{#1}{\textsc{\lowercase{#1}}\xspace}}
\tsc{WGM}
\tsc{QE}
\tsc{EP}
\tsc{PMS}
\tsc{BEC}
\tsc{DE}
\begin{document}
\let\WriteBookmarks\relax
\def\floatpagepagefraction{1}
\def\textpagefraction{.001}

\shorttitle{Emergence of social hierarchies in a society with two competitive groups}    

\shortauthors{Marc Sadurní et~al.}  

\title [mode = title]{Emergence of social hierarchies in a society with two competitive groups} 


%\tnotetext[1]{This document is the results of the research project provided by the ministry’s projects PID2019-106811GB, PRE2020-093266 and PID2022-140757NB-I00. And by Generalitat de Catalunya, grant number 2021SGR00856.}


\author[1,2]{Marc Sadurní}[
    orcid=0000-0002-9870-5513]
    \cormark[1]
%orcid=0000-0002-9870-5513
%\cormark[1]

\ead{marc.sadurni@ub.edu}

\ead[url]{https://www.ubics.net/sadurni-parera-marc/}

\credit{Conceptualization, Methodology, Writing. Original draft}

\affiliation[1]{organization={Departament de Física de la Matèria Condensada, Universitat de Barcelona},
            addressline={Martí i Franquès, 1}, 
            city={Barcelona},
            postcode={08028}, 
            state={Catalonia},
            country={Spain}}
            
\affiliation[2]{organization={Universitat de Barcelona Institute of Complex Systems},
            addressline={Martí i Franquès, 1}, 
            city={Barcelona},
            postcode={08028}, 
            state={Catalonia},
            country={Spain}}

\author[1,2]{Josep Perelló}[
    orcid=0000-0001-8533-6539]
    \cormark[1]

%\cormark[2]
% Footnote of the second author

% Email id of the second author
\ead{josep.perello@ub.edu}

\ead[url]{https://www.ubics.net/perello-palou-josep/}

\credit{Supervision, Writing. Review \& editing}

% Third Author
\author[1,2]{Miquel Montero}[
    orcid=0000-0002-3221-1211]
    \cormark[1]

%\cormark[2]
% Footnote of the third author

% Email id of the third author
\ead{miquel.montero@ub.edu}

% URL of the third author
\ead[url]{https://www.ubics.net/montero-torralbo-miquel/}

\credit{Supervision, Writing. Review \& editing}

\cortext[1]{Corresponding authors}
%\cormark[2]{Principal Corresponding author}

\begin{abstract}
Agent-based models describing social interactions among individuals can help to better understand emerging macroscopic patterns in societies. One of the topics which is worth tackling is the formation of different kinds of hierarchies that emerge in social spaces such as cities. Here we propose a Bonabeau-like model by adding a second group of agents. The fundamental particularity of our model is that only a pairwise interaction between agents of the opposite group is allowed. Agent fitness can thus only change by competition among the two groups, while the total fitness in the society remains constant. The main result is that for a broad range of values of the model parameters, the fitness of the agents of each group show a decay in time except for one or very few agents which capture almost all the fitness in the society. Numerical simulations also reveal a singular shift from egalitarian to hierarchical society for each group. This behaviour depends on the control parameter $\eta$, playing the role of the inverse of the temperature of the system. Results are invariant with regard to the system size, contingent solely on the quantity of agents within each group. Finally, scaling laws are provided thus showing a data collapse from different model parameters and they follow a shape which can be related to the presence of a phase transition in the model.
\end{abstract}

\begin{graphicalabstract}
\includegraphics[width=0.6\textwidth]{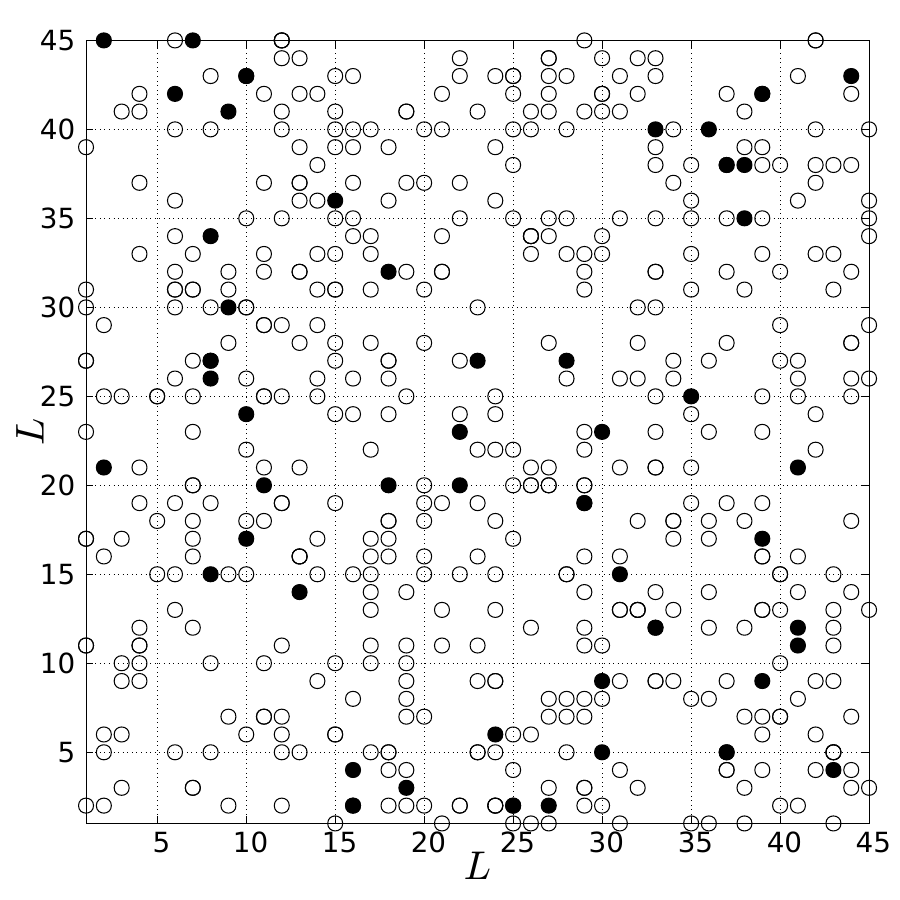}
\end{graphicalabstract}

\begin{highlights}
\item An extended Bonabeau model develops social hierarchies and clear leadership.
\item Dynamics driven by agent reputation interactions is independent of system size.
\item A behavioural transition from egalitarian to hierarchical society is found.
\item Scaling laws are provided to explain the role of the variables of the model.
\end{highlights}

% Keywords

\maketitle

% Main text
%\section{}\label{}

% Use the style of numbering in square brackets.
% If nothing is used, default style will be taken.
%\begin{enumerate}[a)]
%\item 
%\item 
%\item 
%\end{enumerate}  

%\begin{itemize}
%\item 
%\item 
%\item 
%\end{itemize}  

%\begin{description}
%\item[]
%\item[] 
%\item[] 
%\end{description}  

% Figure
%\begin{figure}[<options>]
%	\centering
%		\includegraphics[<options>]{}
%	  \caption{}\label{fig1}
%\end{figure}

%\begin{table}[<options>]
%\caption{}\label{tbl1}
%\begin{tabular*}{\tblwidth}{@{}LL@{}}
%\toprule
%  &  \\ % Table header row
%\midrule
% & \\
% & \\
% & \\
% & \\
%\bottomrule
%\end{tabular*}
%\end{table}
% Uncomment and use as the case may be
%\begin{theorem} 
%\end{theorem}

% Uncomment and use as the case may be
%\begin{lemma} 
%\end{lemma}


%\section{}\label{}
\section{Introduction}\label{sec:introduction}

%%       \citet{<label>} ==> Jones et al. [21]
%%       \cite{<label>} ==> [21]
%%

During last decades, computational science has been playing a crucial role in the study of numerous compelling domains. Within this broad field, agent-based modelling (ABM) simulates behaviour and interactions of heterogeneous units to find regularities and macroscopic properties that can help to better understand real-world phenomena \cite{steinbacher_advances_2021}. In Physics, agent-based modelling has been widely explored in the context of molecular dynamics \cite{alder_1957,alder_1959} with the extensive use of Monte Carlo simulations \cite{metropolis_1953}. Agent-based modelling has also been implemented in the contexts of social physics \cite{jusup_2022} and computational social science (CSS) \cite{lazer_2009}. The main aim in this context is to understand human behaviour within a society in terms of the interaction among their individuals. The basic constituents are not particles but humans interacting with a small number of partners compared to the total size of the system \cite{castellano_statistical_2009}. As a result, statistical physics has been offering valuable insights into various socioeconomic systems. It has contributed to the analysis of macroscopic natural phenomena and it has facilitated the comprehension of large-scale consistencies emerging from the interactions among individual entities \cite{castellano_statistical_2009}. These interactions often produce transitions from disorder to order and the system follows scaling laws that unveil certain universal properties \cite{castellano_statistical_2009}.

Hierarchy formation has been a stimulating complex emergence phenomenon in social sciences \cite{allee_1942,landau_1951,guhl_1968,chase_1980}. In 1951, Landau emphasized that intrinsic characteristics of individuals, such as weight and aggressiveness, proved to be insufficient in explaining the emergence of observed hierarchies and pointed that one shall be carefully looking at the interactions among individuals \cite{landau_1951}. The hierarchy of social organization is an omnipresent property of animal and human aggregations. The emergence of class structures can be observed in many kinds of societies such as insects \cite{wilson_1971}, fishes \cite{goessmann_2000,chase_2002}, birds \cite{lindquist_2009}, mammals \cite{wittig_2003}, and of course in  humans groups \cite{williams_1980,garandeau_2014}. If one particularly consider human societies, then cities become a relevant context in which social inequities organically emerge \cite{moro_mobility_2021,checa_residential_2021,zignani_2019}.

During the 1990s, Bonabeau et al. \cite{bonabeau_phase_1995} proposed a computationally-based model motivated by the empirical observations where dominance relationships seem to be established by the result of fights between individuals. The model relies on the existence of a  feedback mechanism \cite{chase_1994} where individuals who have won more fights are more likely to win future fights. Through a simple algorithm in which individuals move on a square grid, the model thus shows that self-organized hierarchies can emerge spontaneously from an egalitarian society. Many modifications and extensions have been done to make the model more realistic. The most renowned modification was proposed by Stauffer \cite{stauffer_2003b} which involves the adjustment of a parameter within the feedback mechanism of the system. Reference \cite{lacasa_bonabeau_2006} found that the egalitarian solution of the Stauffer version is always stable, while a two-level stable solution (a hierarchical profile) emerges at a critical parameter value via a saddle-node bifurcation. Further improvements of the model are reported in several references \cite{lacasa_bonabeau_2006, malarz_bonabeau_2006, odagaki_self-organizing_2006, tsujiguchi_self-organizing_2007, posfai_talent_2018}. Besides these modifications that aim to enhance realism, there are also articles attempting to simplify the Bonabeau model \cite{ben-naim_dynamics_2005, ben-naim_structure_2006, miyaguchi_piecewise_2020} or even to calibrate the model with real-world data from animals and humans \cite{hickey_self-organization_2019}.

We here introduce a novel approach to the Bonabeau model. We explore a new version of the Bonabeau model where an additional group is introduced in the society. The interactions between individuals of the same group are forbidden, only interactions among individuals of different group are kept. Employing a discrete scheme, we create a mean field approximation that enables us to characterize the intricacies of hierarchical pattern formation. The model can represent the interaction of two distinct groups that might be living or sharing same physical space. The minimalist interacting agent model accounts for the development of social diversity which might be taking place in social contexts such as cities. Interactions might be between two ethnic or cultural populations, between migrants or tourists and native residents. It can also be representing interactions between two polarized income groups, or even between two competing criminal gangs \cite{checa_residential_2021, moro_mobility_2021, zignani_2019, prieto2023reducing}.

The paper is organized as follows. In Section \ref{sec:model}, we introduce an extension of the original Bonabeau model which incorporates a second group into the society and describes the interaction rules between the two groups. In Section \ref{sec:fte}, results via computer simulations are reported. In particular, we describe the time-evolution of key macroscopic values of the model and compute the Gini coefficient under different conditions. In Section \ref{sec:roleofeeta}, we explore the formation of hierarchical structures in the long run. Scaling laws are suggested in Section \ref{sec:USL}. In Section \ref{sec:discussion}, comparisons with the original Bonabeau model and additional hypotheses are discussed. Finally, we conclude in Section \ref{sec:conclusions}, where we summarize key results. We there also include general discussions about the link of the model with real-world phenomena, discuss the limitations of the model and point to possible future work.

\section{The Bonabeau model with two groups of agents}\label{sec:model}

The Bonabeau model \cite{bonabeau_phase_1995} describes the evolution with time $t$ of the so-called fitness functions from $N$ agents that interact within a society. The agents move following random walks on a two-dimensional square lattice $L\times L$. The fitness of an agent $i$ is continuous and changes at every timestep $\Delta t$ due to pairwise interactions of agents sharing the same location in the lattice and also due to a relaxation towards 0 in the absence of any external stimulation. The Bonabeau model understands interactions as a competition between agents. If agent $i$ and agent $j$ interacts, there are two possibilities: (1) agent $i$ wins and therefore the fitness increases by a constant value $\delta^+$ or (2) agent $i$ loses and therefore the fitness decreases by a constant value $\delta^-$. The same rule applies for agent $j$ and its fitness. The outcome of the interaction is assumed to be probabilistic: the larger the value of the fitness, the higher probability of winning the encounter by the agent $i$. The Bonabeau model \cite{bonabeau_phase_1995} assumes for simplicity $\delta^+=\delta^-=1$, so that the fitness is proportional to the number of wins minus the number of loses. 
%In addition to these interactions, a relaxation term that reduces $F_i(t)$ in the absence of any external stimulation is introduced. 
Results of the Bonabeau model show a transition from egalitarian to hierarchical state as the density $\rho=N/L^2$ increases.

In contrast to the original Bonabeau model, our approach assumes the existence of two distinct groups within a society. Agents exclusively compete with agents belonging to the opposing group. We thus designate by $F_{i}^\text{A}(t)$ the fitness of the $N_\text{A}$ agents belonging to a first group $\text{A}$ (where $i\in\{1,\dots,N_\text{A}\}$). And we then designate by $F_{j}^\text{B}(t)$ the fitness of the $N_\text{B}$ agents belonging to a second group $\text{B}$ (where $j\in\{1,\dots,N_\text{B}\}$). At time $t$, agent $i$ with fitness $F_{i}^\text{A}(t)$ interacts with agent $j$ with fitness $F_{j}^\text{B}(t)$ if they both occupy the same location. After the interaction, they exchange a certain amount of fitness proportional to parameter $0<x<1$. If agent $i$ wins, the fitnesses change during time step $\Delta t$ in the following way:
\begin{equation}
\begin{aligned}
F_{i}^\text{A}(t+\Delta t) &=& F_{i}^\text{A}(t)+x F_{j}^\text{B}(t),\\
F_{j}^\text{B}(t+\Delta t) &=& F_{j}^\text{B}(t)-x F_{j}^\text{B}(t),
\end{aligned}
\label{eq:fitnessevolution1}
\end{equation}
while if agent $i$ loses, the fitnesses exchange read:
\begin{equation}
\begin{aligned}
F_{i}^\text{A}(t+\Delta t) &=& F_{i}^\text{A}(t)-x F_{i}^\text{A}(t),\\
F_{j}^\text{B}(t+\Delta t) &=& F_{j}^\text{B}(t)+x F_{i}^\text{A}(t).
\end{aligned}
\label{eq:fitnessevolution2}
\end{equation}
Unlike the original Bonabeau model, here we do not include a relaxation term \cite{bonabeau_phase_1995}. 

Winning or losing depends on probability $P_{i j}(t)$. Following the definition from the original Bonabeau model \cite{bonabeau_phase_1995}, winning probability of agent $i$ over agent $j$ reads: 
\begin{equation}
P_{i j}(t)=\frac{1}{1+\exp \left[\eta\left(\hat{F}_{j}^\text{B}(t)-\hat{F}_{i}^\text{A}(t)\right)\right]},
\label{eq:probbusiness}
\end{equation}
where we have defined the normalized fitness as:
\begin{equation}
\begin{aligned}
\hat{F}_{i}^\text{A}(t)=\frac{F_{i}^\text{A}(t)-F_\text{min}^\text{A}(t)}{F_\text{max}^\text{A}(t)-F_\text{min}^\text{A}(t)}, \\
\hat{F}_{j}^\text{B}(t)=\frac{{F}_{j}^\text{B}(t)-{F}_\text{min}^\text{B}(t)}{{F}_\text{max}^\text{B}(t)-{F}_\text{min}^\text{B}(t)},
\end{aligned}
\label{eq:normalization1}
\end{equation}
where $F_\text{min}^\text{A}(t)\equiv \mbox{min}\{F_{i}^\text{A}(t)\}$ and $F_\text{min}^\text{B}(t)\equiv \mbox{min}\{F_{j}^\text{B}(t)\}$ are the smallest non-normalized fitness values for each group, and $F_\text{max}^\text{A}(t)=\mbox{max}\{F_{i}^\text{A}(t)\}$ and $F_\text{max}^\text{B}(t)=\mbox{max}\{F_{j}^\text{B}(t)\}$ are the highest non-normalized fitness values for each group. Two alternative ways of normalizing fitness are presented in \textit{Appendix \ref{sec:Norm}}. They both show the same qualitative behaviour.

Equation (\ref{eq:probbusiness}) also incorporates the parameter $\eta > 0$ which regulates the intensity of the interactions. In physical terms, $\eta$ can be interpreted as the inverse of the temperature of the system \cite{bonabeau_phase_1995}. The winning probability $P_{i j}(t)$ depends on the normalized fitness difference of agents $j$ and $i$. The larger the normalized fitness of the agent $i$, the higher the probability of winning. Note that if $\hat{F}_{i}^\text{A}(t)\gg \hat{F}_{j}^\text{B}(t)$, $P_{i j}(t)$ tends to 1. If $\hat{F}_{i}^\text{A}(t)\ll \hat{F}_{j}^\text{B}(t)$, $P_{i j}(t)$ tends to 0.

We can also define the overall fitness for the entire society as:
\begin{equation}
\phi= \sum_{i=1}^{N_\text{A}} F_{i}^\text{A}(t)+\sum_{j=1}^{N_\text{B}} F_{j}^\text{B}(t) = F_{\text{tot}}^\text{A}(t) + F_{\text{tot}}^\text{B}(t).
\label{eq:total}
\end{equation}
The total fitness $\phi$ is constant (independent of time $t$) because the interaction between two agents always preserve the total fitness.

All agents of both groups are moving on a $L\times L$ square lattice following a random walk (for one static group, the results qualitatively do also not vary). In order to reproduce the time that a jump happens and also to determine which agent group moves, a residence time algorithm (the Gillespie algorithm) has been implemented \cite{raul}. Agents are bosons, i.e., there can be many agents in the same site. The interaction takes place when the moving agent goes to a site that is occupied by one or more agents of the opposite group. If there are multiple agents from the other group, one single agent is randomly chosen to interact with. A flux diagram of the code is shown in Figure \ref{f:fluxdiagram}. More details of the simulation techniques are reported in \textit{Appendix \ref{sec:MC}} and the computer code is available in GitHub \cite{github}.

\begin{figure*}[btp!]
    \includegraphics[width=\textwidth,height=10cm]{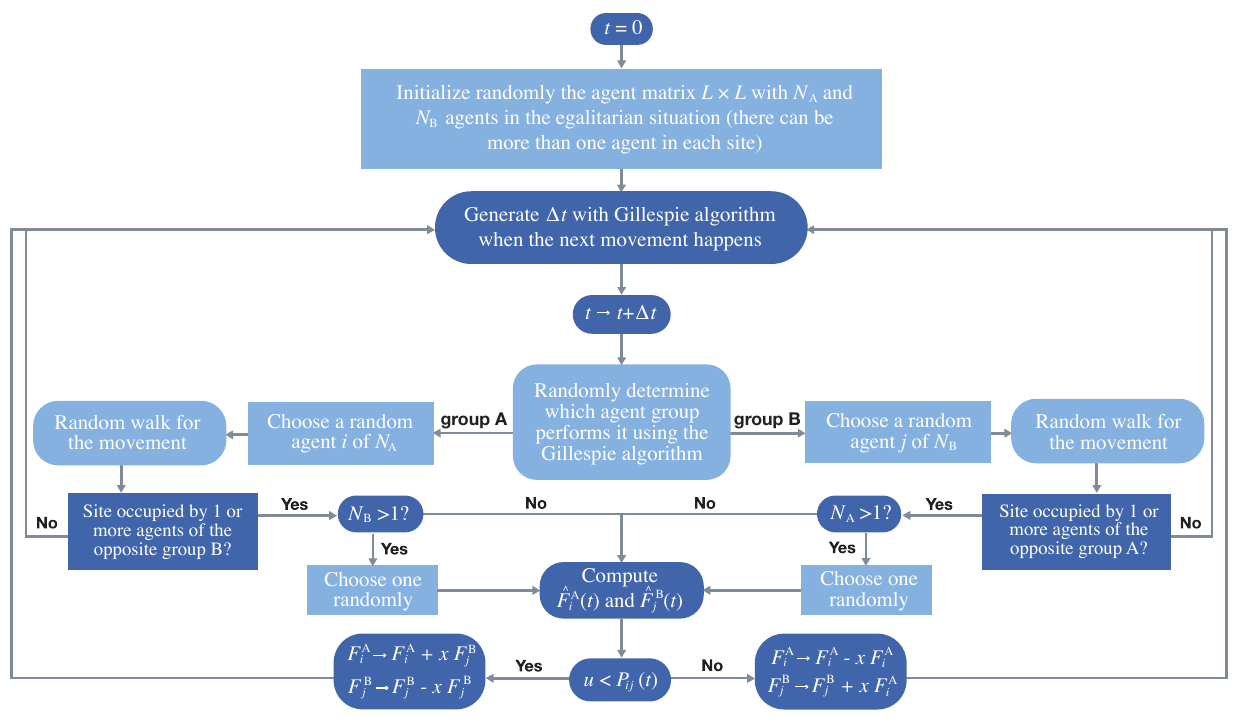}
  \caption{\textbf{Flux code diagram of computer simulations.} Flux code diagram of the Monte Carlo simulations with the Gillespie algorithm implemented, where $u\sim U(0,1)$ is a uniform random number between 0 and 1. A detailed description can be found in \textit{Appendix \ref{sec:MC}}.}
  \label{f:fluxdiagram}
\end{figure*}

\section{Fitness temporal evolution}\label{sec:fte}

\begin{figure}[btp!]
	\centering
		\includegraphics[scale=.50]{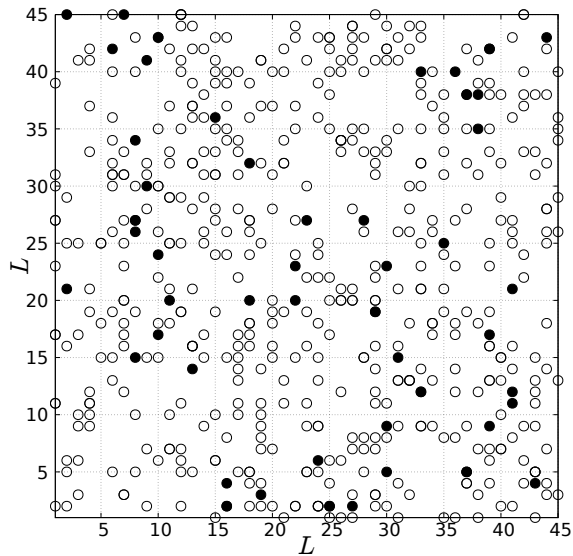}
		\caption{\textbf{Initial spatial distribution of agents in a regular square lattice.} Initial random positions of both groups of agents with $N_\text{A}=500$ (empty circles) and $N_\text{B}=50$ (filled circles) for $L=45$. There can be more than one agent in each site.}
	\label{f:Inipositions}
\end{figure}

Initially (at time $t=0$), all agents are randomly distributed over the lattice. See in Figure \ref{f:Inipositions} one initial setting to exemplify agents' distribution in a $L\times L$ square lattice. Their fitness value is initially set to $F_{i}^\text{A}(0)=F_{j}^\text{B}(0)=\phi/({N_\text{A}+N_\text{B}})$: the so-called egalitarian regime. As defined, the results are entirely invariant to the total fitness of the whole society. For clarity, we set $\phi$ to $1\,000$ arbitrary units for all simulations presented in this section.

\subsection{Parameter dependence} \label{sec:parameters}

\begin{figure}[btp!]
	\centering
        \begin{subfigure}[b]{0.45\textwidth}
		\includegraphics[width=\textwidth]{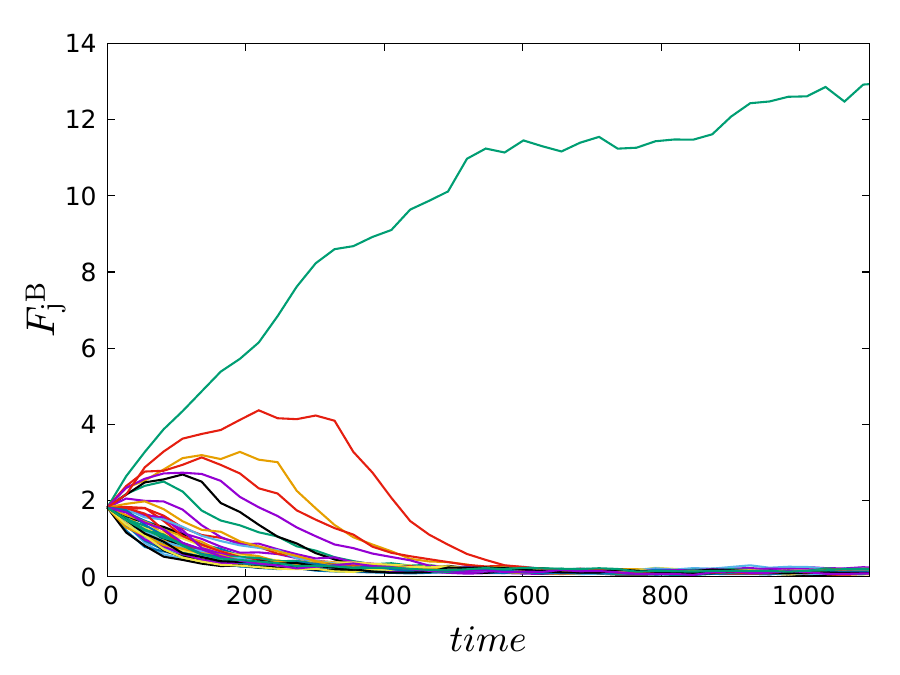}
              \caption{}
        \end{subfigure}
        \begin{subfigure}[b]{0.45\textwidth}
		\includegraphics[width=\textwidth]{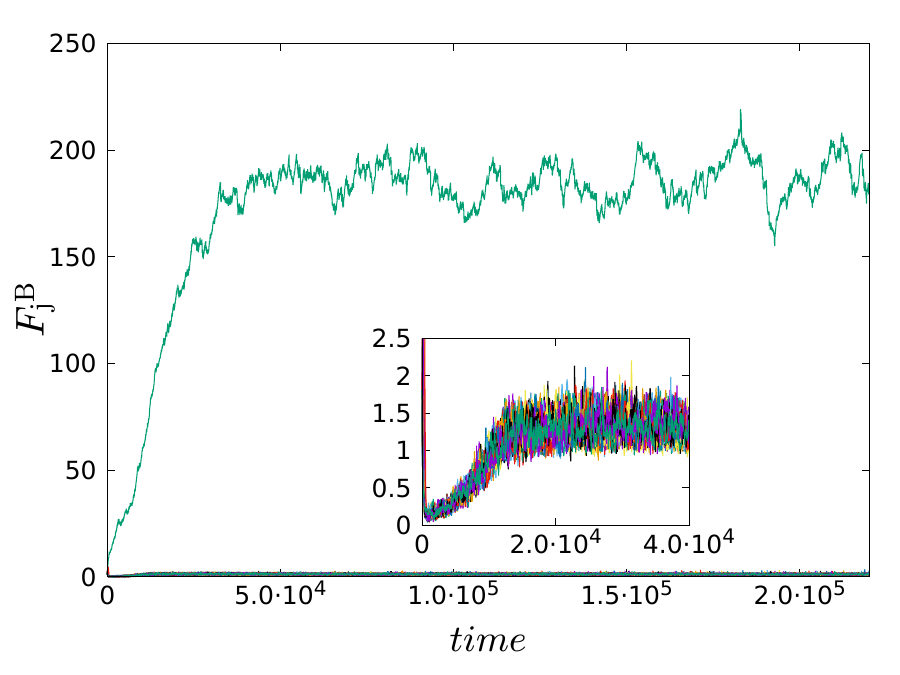}
              \caption{}
        \end{subfigure}
	\caption{\textbf{Temporal evolution of the fitnesses of group B.} Time evolution of all $F_{j}^\text{B}$ for $N_\text{A}=500$ and $N_\text{B}=50$ individuals, $\eta= 5$ and $x=0.01$, randomly simulated on a $L\times L$ lattice ($L=25$). (a) The first $1\,100$ time steps. (b) Larger time window to observe the transient and stationary fluctuations of the leader while the inset shows the behaviour of the other agents of the group.}
	\label{f:5-001-L25}
\end{figure}

We first conduct a coarse-grained analysis of the temporal evolution of agents' fitness to examine its reliance on all model parameters. Figure \ref{f:5-001-L25} shows the time evolution of the fitness of all agents from the group B ($N_\text{B}=50$). We there fix a particular set of values of the parameters and then represent the first $1\,100$ steps in Figure \ref{f:5-001-L25} (a) to see how a leader emerges in this specific model setup. The same applies to the group A. Figure \ref{f:5-001-L25} (b) shows how the leader fluctuates around a stationary value for large enough times (the stationary regime). The inset of Figure \ref{f:5-001-L25} (b) illustrates the transient behaviour for the rest of agents (those keeping just a small portion of the total fitness). The phenomena shown in the inset arises when a single leader emerges in the opposite group, triggering subsequent fitness exchanges among the rest of the agents until a stable regime is reached.

\begin{figure}[btp!]
	\centering
        \begin{subfigure}[b]{0.45\textwidth}
		\includegraphics[width=\textwidth]{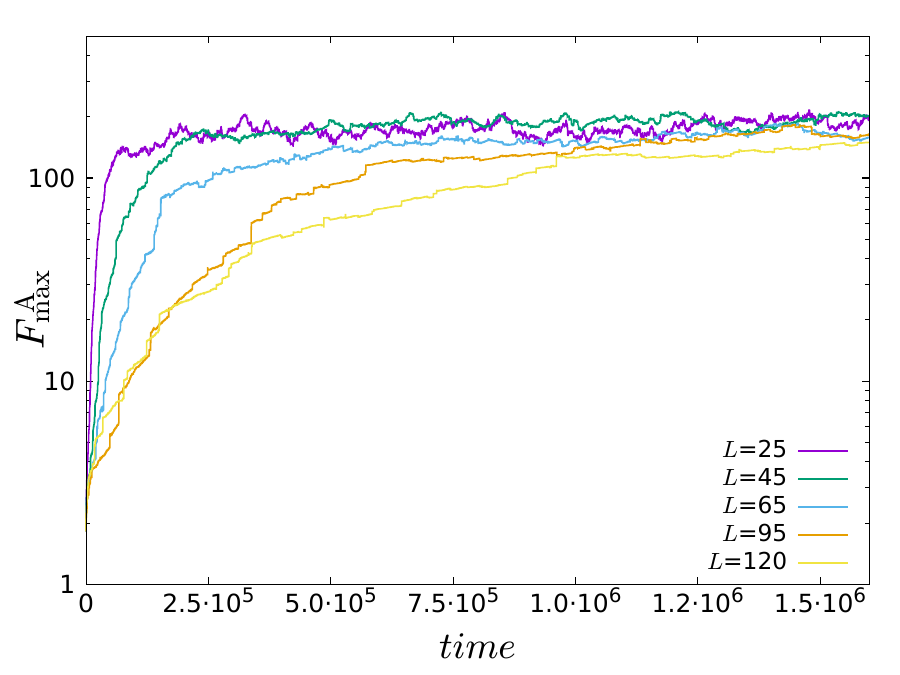}
              \caption{}
        \end{subfigure}
        \begin{subfigure}[b]{0.45\textwidth}
		\includegraphics[width=\textwidth]{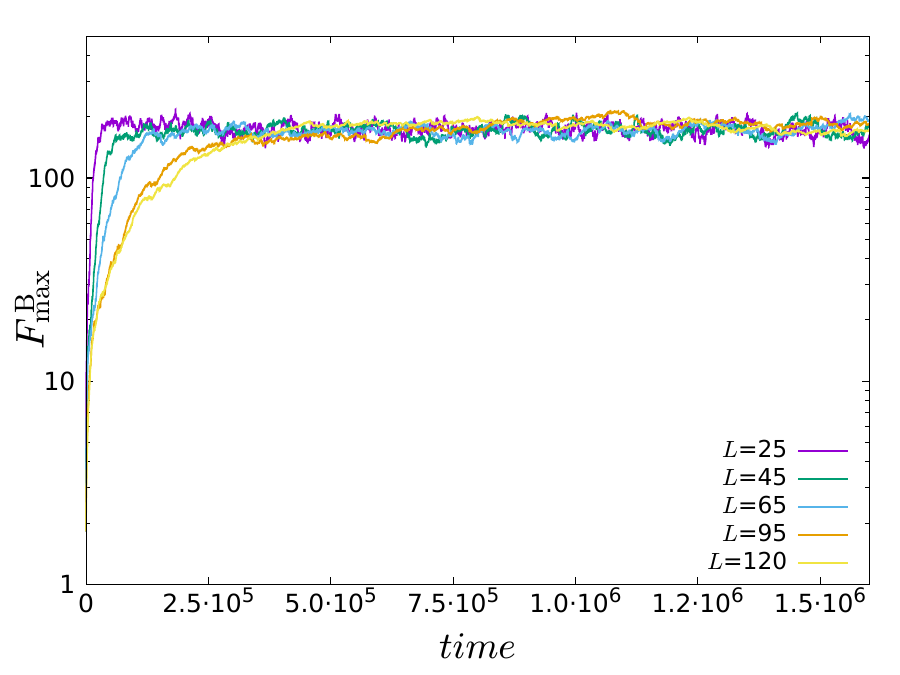}
              \caption{}
        \end{subfigure}
		\caption{\textbf{Maximum fitness temporal evolution of the two groups for several systems sizes.} Time evolution of the maximum fitness for $N_\text{A}=500$ and $N_\text{B}=50$ agents, $\eta=5$ and $x=0.01$, randomly simulated on several $L\times L$ square lattice sizes (largest to smallest ascending). (a) $F_\text{max}^\text{A}(t)$ and (b) $F_\text{max}^\text{B}(t)$. The vertical axis is in log scale.}
	\label{f:fmax-s}
\end{figure}

After repeating the simulations while only modifying the system size $L$, we get the trajectories showed in \textit{Appendix \ref{sec:Systemsize}}. We can observe that, regardless of the value of $L$, one agent emerges as a leader for the same parameters in Figure \ref{f:5-001-L25}. We will later see that the leader emergence might not be always the case as this will depend on the parameters choice. Figure \ref{f:fmax-s} closely examines both $F_\text{max}^\text{A}(t)$ and $F_\text{max}^\text{B}(t)$ in the stationary regime. We there explore several system sizes $L$. The time required to reach the stationary regime is longer for larger values of $L$. Indeed, we observe that the time required to reach the stationary regime for maximum fitness increases in a polynomial manner with respect to the system size $L$ as described in \textit{Appendix \ref{sec:Systemsize}}. Except for illustration purposes, we keep the system size constant setting the $L\times L$ square lattice to a constant value $L=25$. This relatively small value leads to a quicker attainment of the stationary regime and it reduces CPU time.

\begin{figure*}[btp!]
	\centering
        \begin{subfigure}[b]{0.45\textwidth}
		\includegraphics[width=\textwidth]{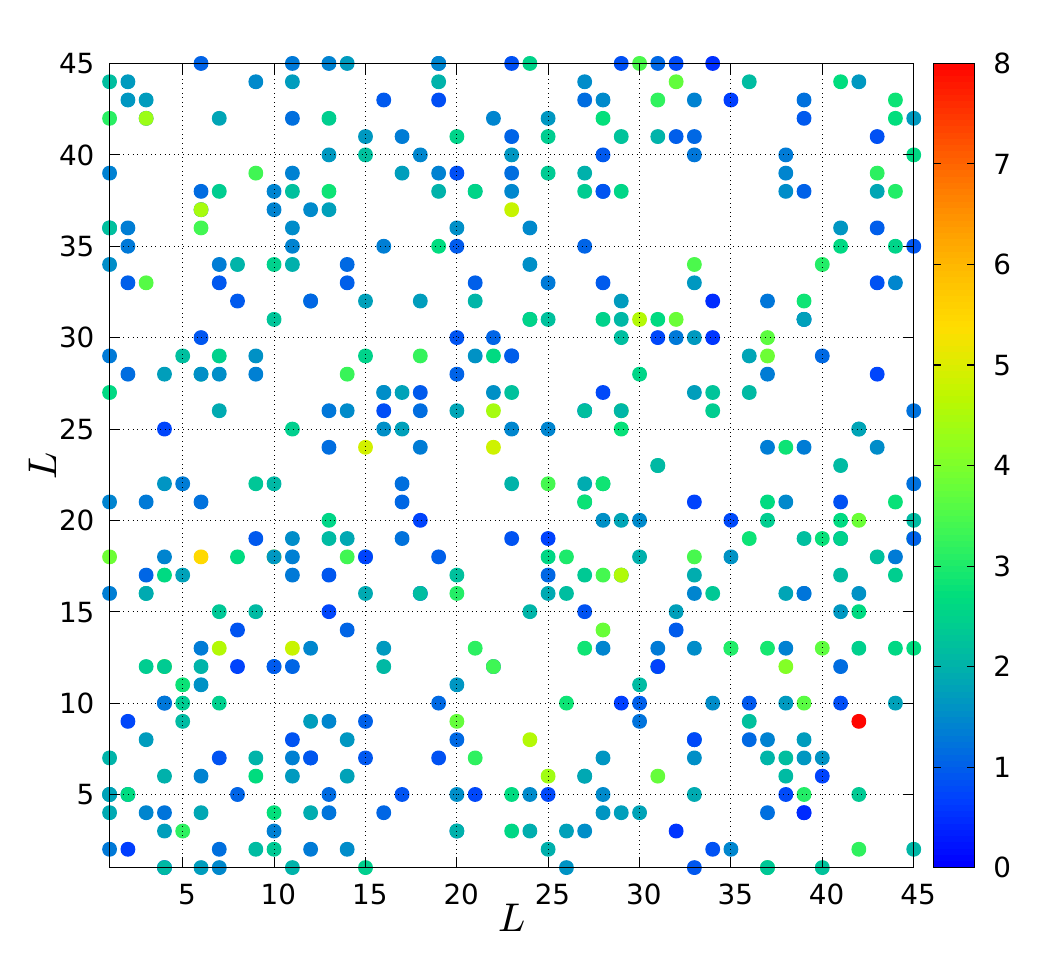}
            \caption{}
        \end{subfigure}
        \begin{subfigure}[b]{0.45\textwidth}
		\includegraphics[width=\textwidth]{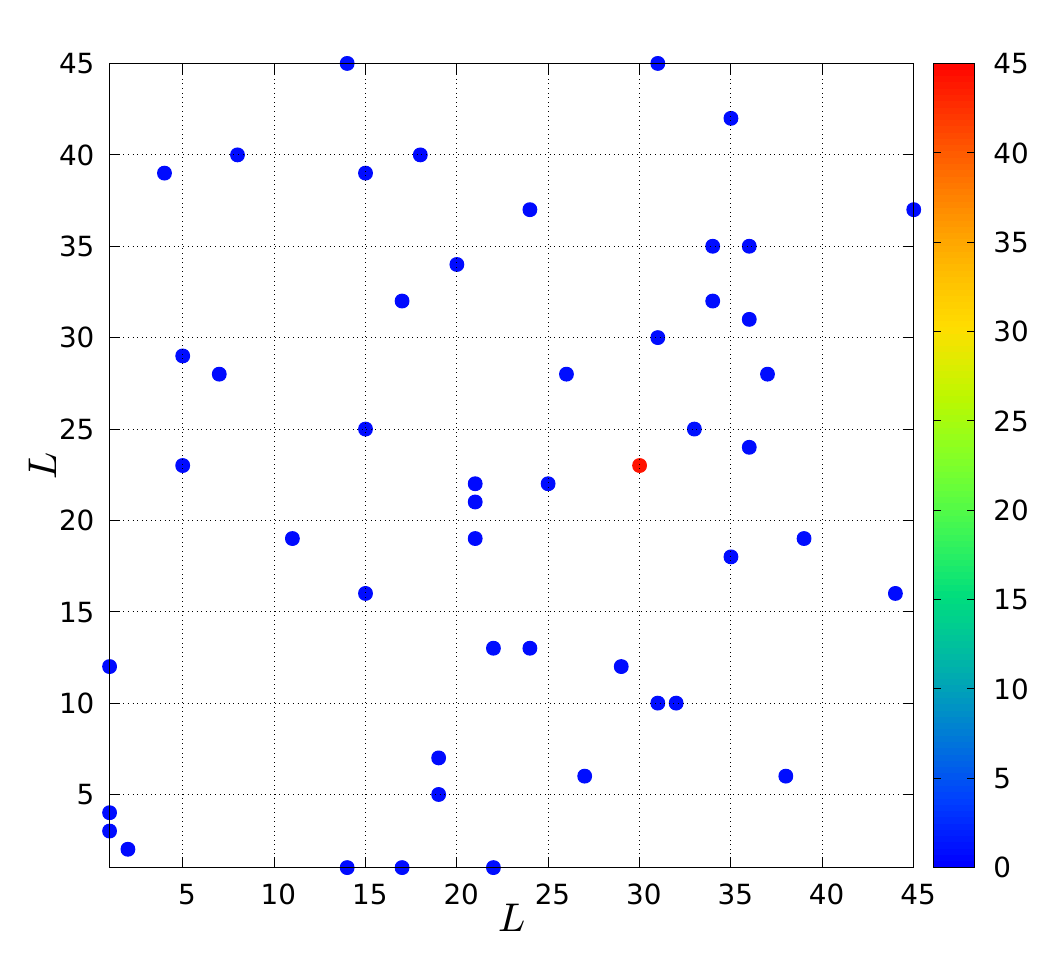}
            \caption{}
        \end{subfigure}
        \begin{subfigure}[b]{0.45\textwidth}
		\includegraphics[width=\textwidth]{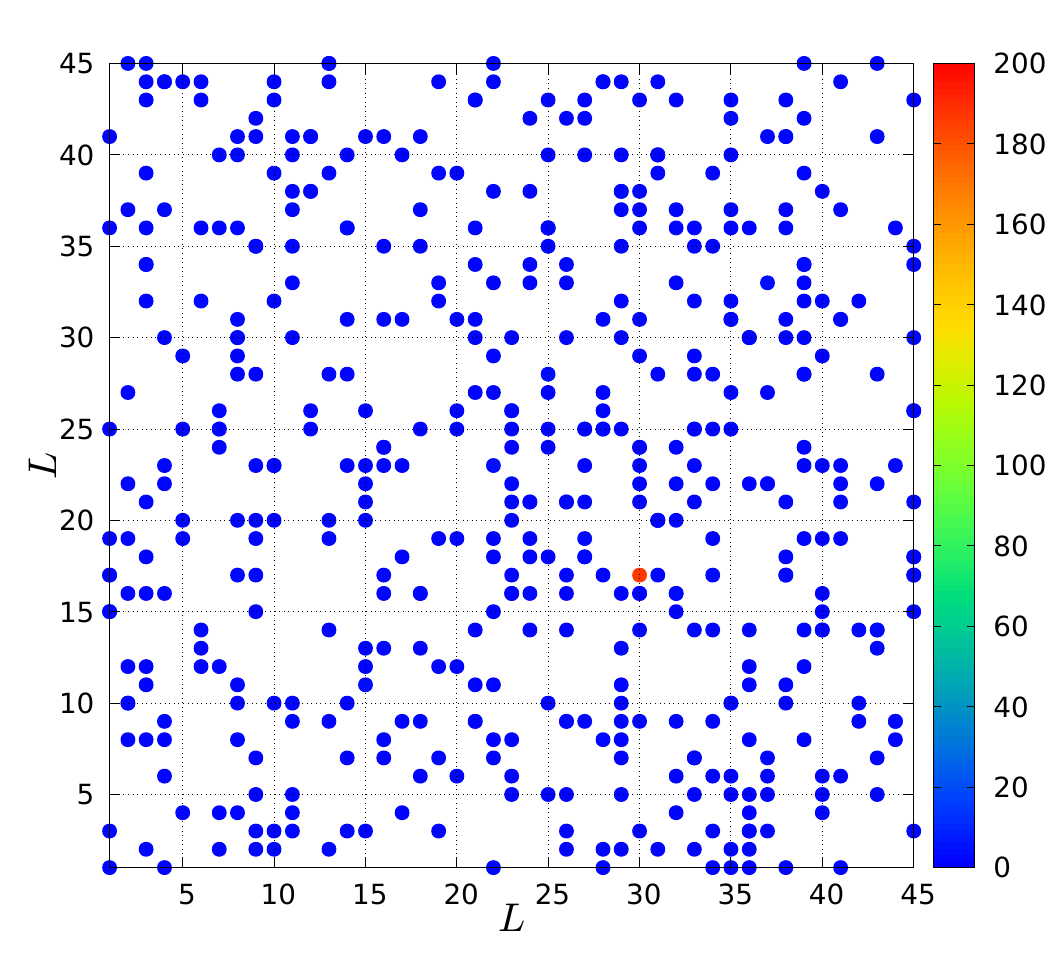}
            \caption{}
        \end{subfigure}
        \begin{subfigure}[b]{0.45\textwidth}
		\includegraphics[width=\textwidth]{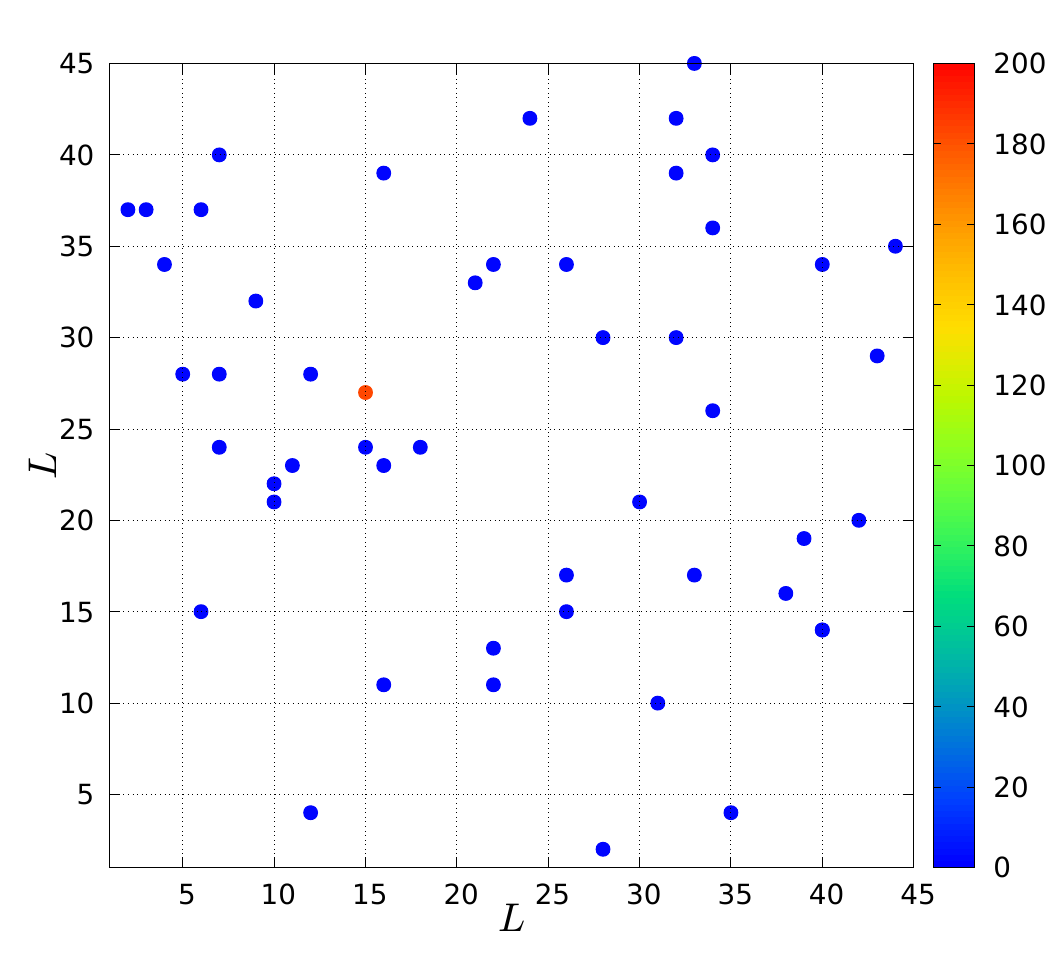}
            \caption{}
        \end{subfigure}
	\caption{\textbf{Fitness distribution mapping for the two groups at two distinct times.} Fitness heat map of both groups of a single simulation for $N_\text{A}=500$ agents, $N_\text{B}=50$ agents, $\eta= 5$, $x=0.01$ in a $L\times L$ square lattice with $L=45$. (a) $20\,000$ Gillespie time units and group A. (b) 20\,000 Gillespie time units and group B. (c) Stationary regime at $600\,000$ Gillespie time units and group A. (d) Stationary regime at $600\,000$ Gillespie time units and group B. There can be more than one agent in each site.}
	\label{f:spatialfitnessbothclasses}
\end{figure*}

We now compare the time evolution of the two groups with $N_\text{A}$ and $N_\text{B}$ agents in Figure \ref{f:spatialfitnessbothclasses}. Figure \ref{f:spatialfitnessbothclasses} (b) shows that after $20\,000$ Gillespie time units, the group B ($N_\text{B}\ll N_\text{A}$) is dominated by a single agent. In contrast, Figure \ref{f:spatialfitnessbothclasses} (a) shows that there are still multiple agents sharing an equal amount of fitness in the group A. No marked leader has yet emerged. Therefore, inequalities emerge more rapidly in the group with the smallest size.

\begin{figure*}[btp!]
	\centering
        \begin{subfigure}[b]{0.45\textwidth}
		\includegraphics[width=\textwidth]{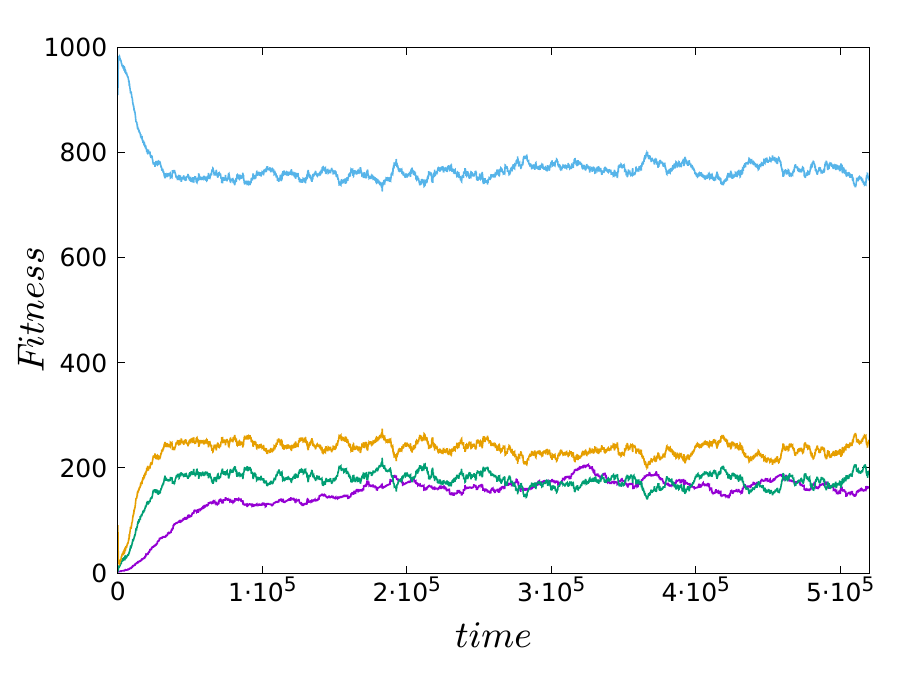}
            \caption{}
        \end{subfigure}
        \begin{subfigure}[b]{0.45\textwidth}
		\includegraphics[width=\textwidth]{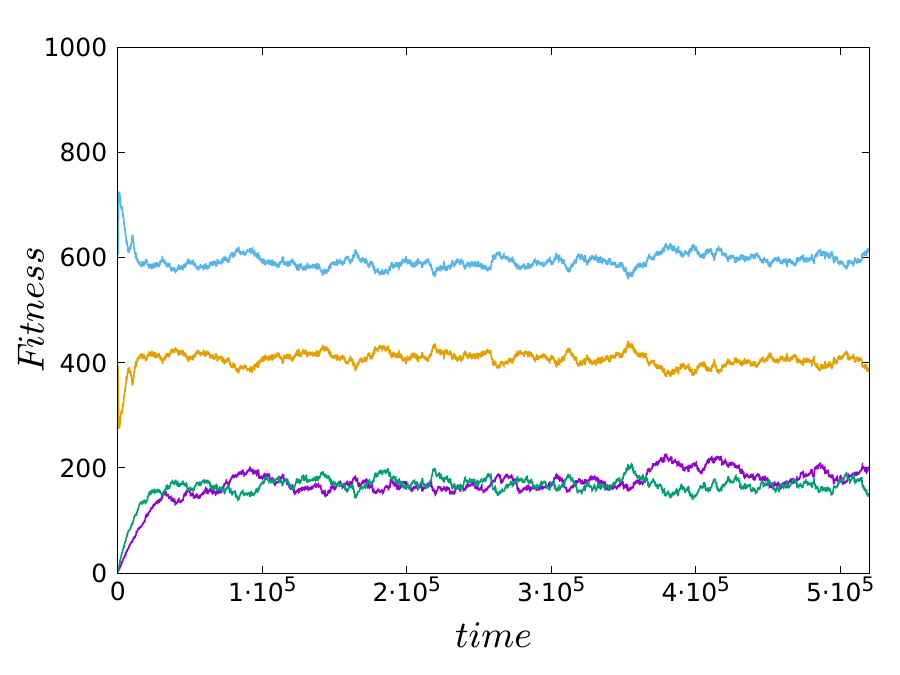}
            \caption{}
        \end{subfigure}
        \begin{subfigure}[b]{0.45\textwidth}
		\includegraphics[width=\textwidth]{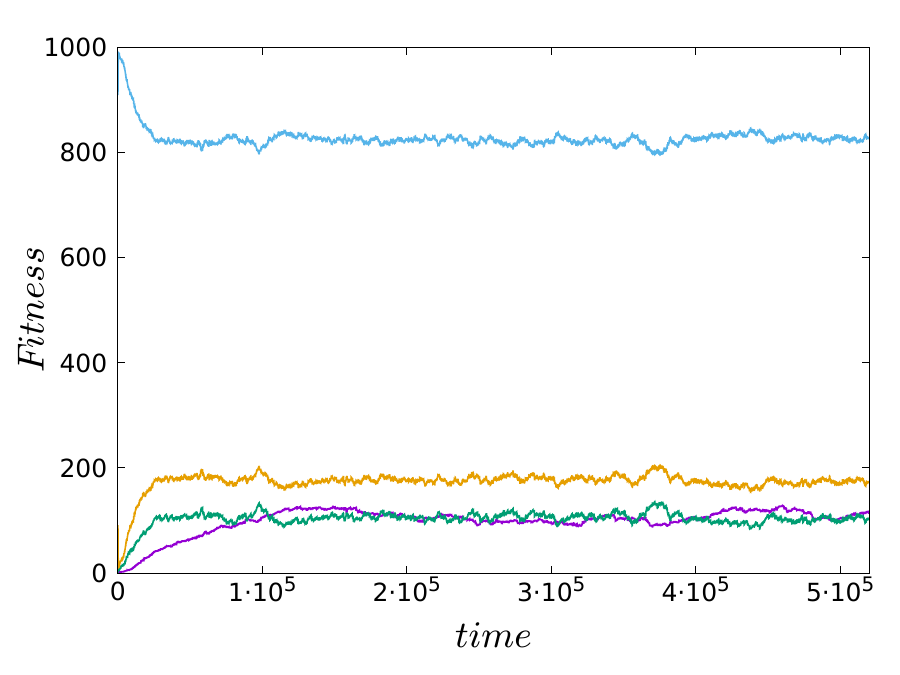}
            \caption{}
        \end{subfigure}
        \begin{subfigure}[b]{0.45\textwidth}
		\includegraphics[width=\textwidth]{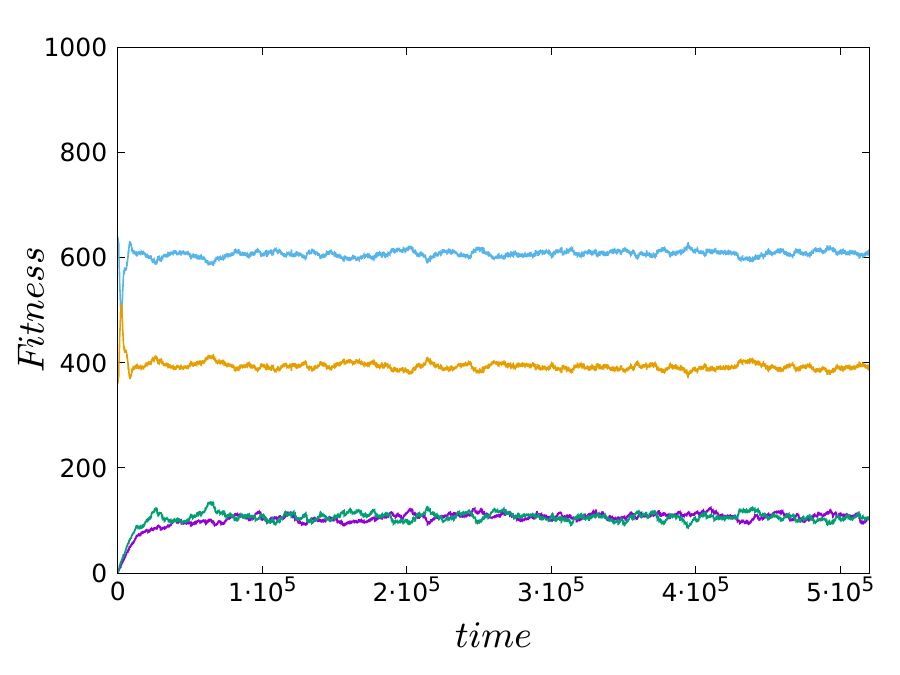}
            \caption{}
        \end{subfigure}
	\caption{\textbf{Temporal evolution of key indicators depending on the number of agents.} Time evolution of $F_\text{max}^\text{A}(t)$ (purple), $F_\text{max}^\text{B}(t)$ (green), $F_\text{tot}^\text{A}(t)$ (blue) and $F_\text{tot}^\text{B}(t)$ (yellow) for $\eta=5$ and $x=0.01$, randomly simulated on a $L\times L$ lattice ($L=25$). (a) $N_\text{A}=500$ and $N_\text{B}=50$. (b) $N_\text{A}=350$ and $N_\text{B}=200$. (c) $N_\text{A}=1\,000$ and $N_\text{B}=100$. (d) $N_\text{A}=700$ and $N_\text{B}=400$.}
	\label{f:N-L25-o}
\end{figure*}

We can also vary the number of agents in the system ($N_\text{A}$ and $N_\text{B}$). In Figure \ref{f:N-L25-o} we consider the maximum fitness values of each group, $F_\text{max}^\text{A}(t)$ and $F_\text{max}^\text{B}(t)$, and their total fitness $F_\text{tot}^\text{A}(t)$ and $F_\text{tot}^\text{B}(t)$ for several combinations of the number of agents. In particular, in Figure \ref{f:N-L25-o} (c) compared to Figure \ref{f:N-L25-o} (a), the fraction $N_\text{A}/N_\text{B}=10$ is kept constant by doubling $N_\text{A}$ and $N_\text{B}$. On the other hand, in Figure \ref{f:N-L25-o} (b) the total number of agents $N_\text{T}=N_\text{A}+N_\text{B}=550$ is maintained constant in respect to Figure \ref{f:N-L25-o} (a) while the fraction $N_\text{A}/N_\text{B}=1.75$ is reduced, making groups more equal in size. Ultimately, the number of agents illustrated in Figure \ref{f:N-L25-o} (d) can be determined from Figure \ref{f:N-L25-o} (b) by keeping constant the fraction $N_\text{A}/N_\text{B}=1.75$ and doubling $N_\text{A}$ and $N_\text{B}$. Analysing in detail the four Figures, several conclusions can be drawn. In the first case, while maintaining the fraction constant, and duplicating $N_\text{T}$, results in impact on $F_\text{tot}^\text{A}(t)$ and $F_\text{tot}^\text{B}$, and also notable reductions in $F_\text{max}^\text{A}(t)$ and $F_\text{max}^\text{B}(t)$ as $N_\text{T}$ increases. Conversely, when the total number of agents $N_\text{T}$ remains the same but the ratio $N_\text{A}/N_\text{B}$ changes, a quite different behaviour is observed. The stationary regime for the maximum fitnesses remain invariable, but there is a considerable alteration in the distribution of fitness between groups. Indeed, the emergence of a leader is closely tied to the distance that separates the values of $F_\text{max}^\text{A}(t)$ and $F_\text{tot}^\text{A}(t)$ and the values of $F_\text{max}^\text{B}(t)$ and $F_\text{tot}^\text{B}(t)$. An increased gap between these two lines corresponds to a reduced inequality among agents in a group, while a narrower gap signifies greater inequality among agents. Hence, the stationary regime for the maximum fitness and the sum of fitness within each group is highly contingent on $N_\text{A}$, $N_\text{B}$ and their ratio $N_\text{A}/N_\text{B}$. Indeed, as $N_\text{A}/N_\text{B}$ decrease with fixed $N_\text{T}$, the gap is amplified for the minority group, resulting in reduced inequalities within that particular group. However, the gap diminishes for the majority group, giving rise to greater disparities. Our model exhibits a high level of sensitivity to the interactions between agents from opposing groups. If the ratio between the number of agents in both groups shifts, it affects not only the number of interactions but also the fitness values associated with those exchanges.

\begin{figure}[btp!]
	\centering
         \begin{subfigure}[b]{0.45\textwidth}
		\includegraphics[width=\textwidth]{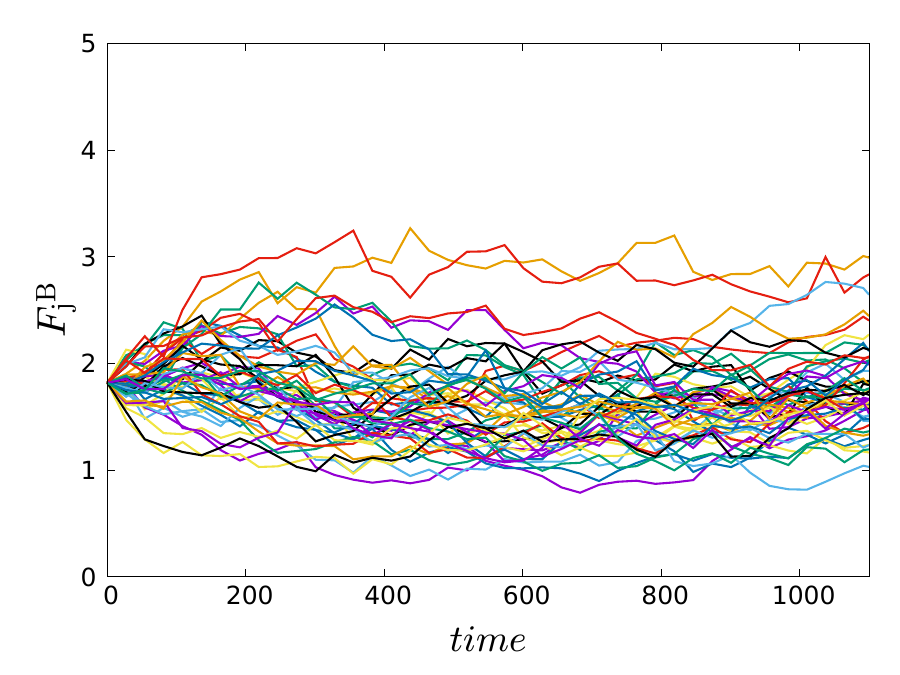}
            \caption{}
        \end{subfigure}
        \begin{subfigure}[b]{0.45\textwidth}
		\includegraphics[width=\textwidth]{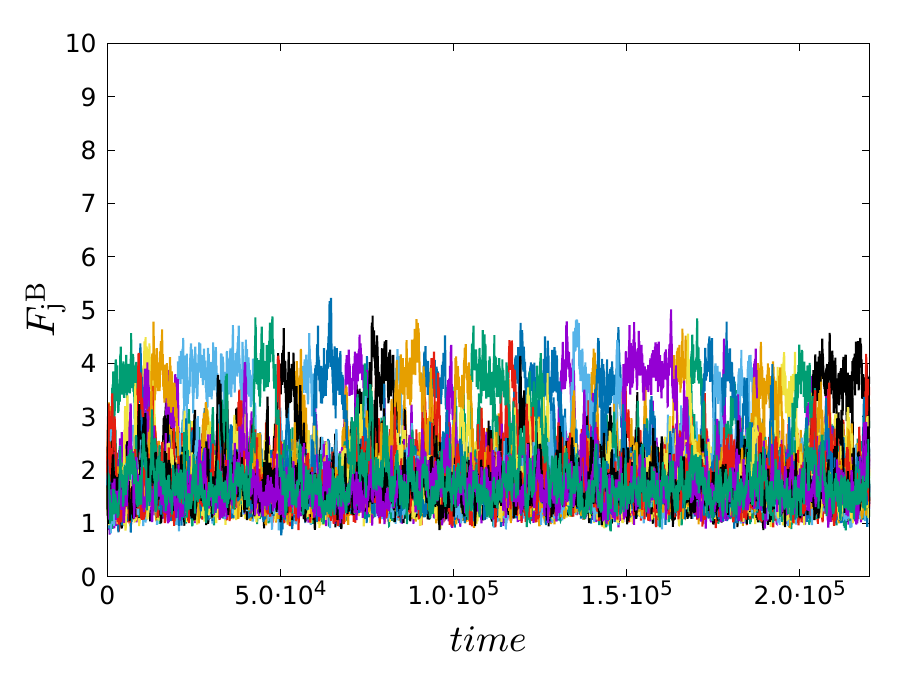}
            \caption{}
        \end{subfigure}
	\caption{\textbf{Fitness temporal evolution of one group for smaller $\eta$.} Time evolution of all $F_{j}^\text{B}$ for $N_\text{A}=500$ and $N_\text{B}=50$ agents, $\eta=1$ and $x=0.01$, randomly simulated on a $L\times L$ lattice ($L=25$). (a) During the first $1\,100$ time steps. (b) Stationary regime.}
	\label{f:1-001-L25}
\end{figure}

\begin{figure}[btp!]
	\centering
        \begin{subfigure}[b]{0.45\textwidth}
		\includegraphics[width=\textwidth]{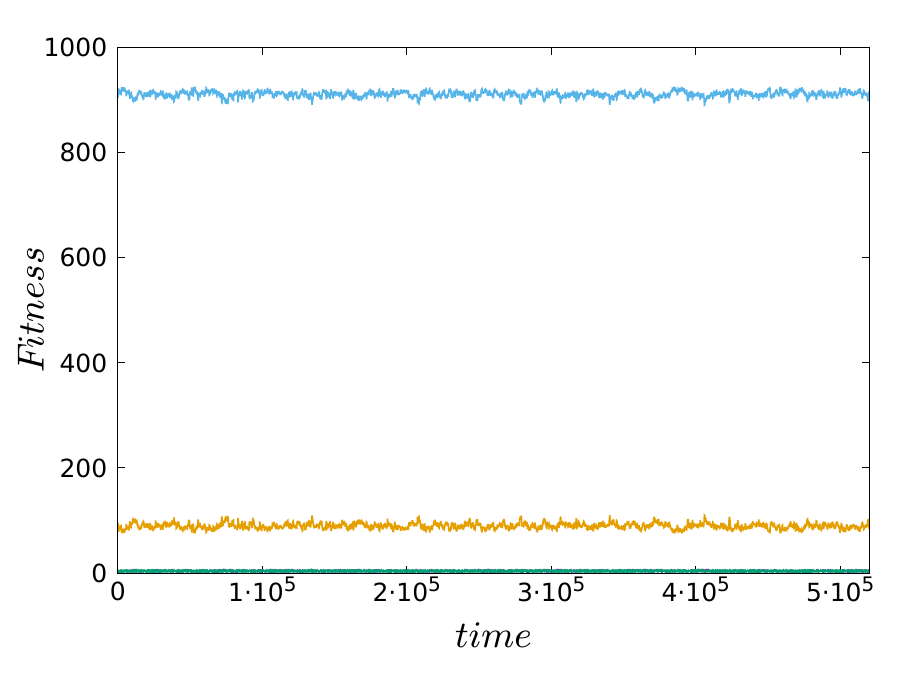}
            \caption{}
        \end{subfigure}
        \begin{subfigure}[b]{0.45\textwidth}
		\includegraphics[width=\textwidth]{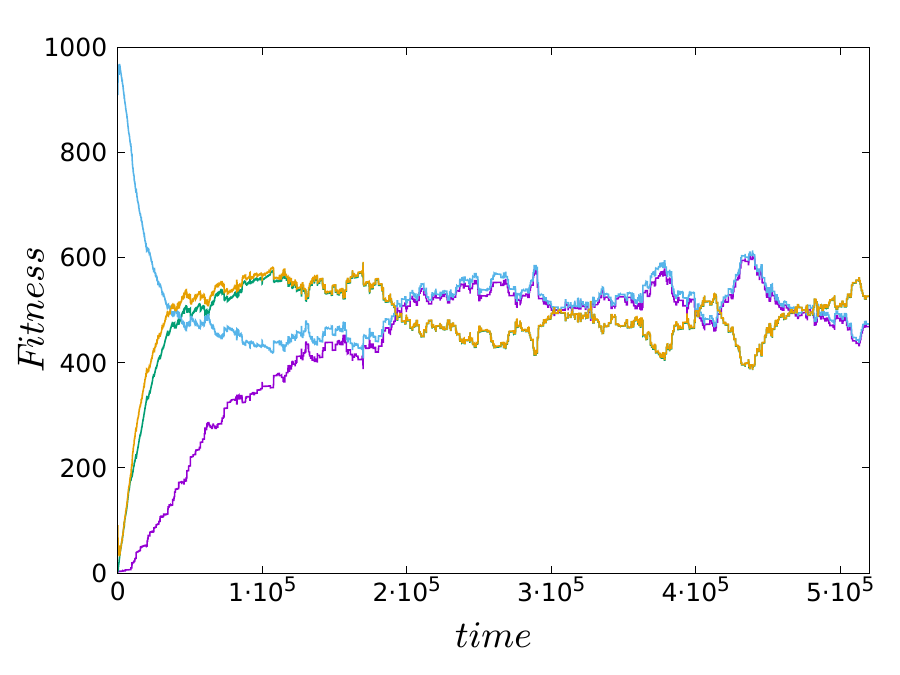}
            \caption{}
        \end{subfigure}
	\caption{\textbf{Temporal evolution of key indicators with two different $\eta$ values.} Time evolution of $F_\text{max}^\text{A}(t)$ (purple), $F_\text{max}^\text{B}(t)$ (green), $F_\text{tot}^\text{A}(t)$ (blue) and $F_\text{tot}^\text{B}(t)$ (yellow) for $N_\text{A}=500$ and $N_\text{B}=50$ agents and $x=0.01$, randomly simulated on a $L\times L$ lattice ($L=25$). (a) $\eta=1$. (b) $\eta=10$.}
	\label{f:5-001-L-o}
\end{figure}

Figure \ref{f:1-001-L25} modifies $\eta$. We set a smaller value compared to previous simulations, so that, there is no clear leader in the long run. This result indicates that the model behaves differently depending on the $\eta$ parameter: we obtain an egalitarian or a hierarchical society depending on $\eta$. Figure \ref{f:5-001-L-o} takes a closer look to this phenomena by considering again the four key indicators in our model: the maximum fitness values of each group, $F_\text{max}^\text{A}(t)$ and $F_\text{max}^\text{B}(t)$, and the total fitness for each of one $F_\text{tot}^\text{A}(t)$ and $F_\text{tot}^\text{B}(t)$. Results show that $\eta$ significantly influences the evolution of the indicators. If $\eta$ is higher, inequalities are magnified, ultimately culminating in the emergence of a single leader within each group of the society. 

The exchange factor of fitness, denoted as $x$, does not impact on the outcomes described. As shown in \textit{Appendix \ref{sec:exchangefactor}}, the fitness is however much more stochastic and the time that the system spends to reach a unique leader under certain conditions also varies. If the proportion exchanged is large enough, an agent with small portion of fitness could grow and can lead to a drastic decrease of the fitness of the leader of the opposite group, and so on cyclically. Altering the number of agents in each group and $\eta$ parameter, while keeping the system size constant, leads to distinct outcomes. These outcomes are further elaborated in the subsequent sections.

To conclude this section, it can be stated that the emergence of a leader or a limited number of leaders in the model is primarily contingent on the values of $\eta$ and the number of agents ($N_\text{A}$ and $N_\text{B}$) and, in turn, the total number of agents $N_\text{T}$, but also on the ratio $N_\text{A}/N_\text{B}$. The model describes a shift from an egalitarian to a hierarchical society with changes on $\eta$, $N_\text{A}$, $N_\text{B}$ and $N_\text{A}/N_\text{B}$. On the contrary, the shift does not seem to be caused by changes on the system size $L$ and on the exchange factor $x$. Table \ref{tab:parameters} summarizes how the response of the system changes with respect to the mentioned model parameters.

\begin{table*}[btp!]
\caption{\textbf{Dependence on model parameters.} The effect of each model parameter on the stationary fitness values and on the time required to reach the stationary regime (“$\checkmark$” means “it depends” and “$-$” means “it does not depend”).}
\centering
\begin{tabular}{cccc}
\hline
\begin{tabular}{c}
Model \\ parameters
\end{tabular} & \begin{tabular}{c}
Definition
\end{tabular} & \begin{tabular}{c}
Stationary \\ fitness
\end{tabular} & \begin{tabular}{c} 
Time to \\ stationarity
\end{tabular} \\
\hline $L$ & System size & $-$ & $\checkmark$ \\
$\eta$ & Interaction intensity & $\checkmark$ & $\checkmark$ \\
$x$ & Exchange factor & $-$ & $\checkmark$ \\
$N_\text{A}$ $\&$ $N_\text{B}$ & Number of agents in group A and B & $\checkmark$ & $\checkmark$ \\
$N_\text{T}$ & Total number of agents in the society & $\checkmark$ & $\checkmark$ \\
$\phi$ & Overall fitness of the entire society & $-$ & $-$ \\
$F_{i}^\text{A}(0)$ $\&$ $F_{j}^\text{B}(0)$ & Initial fitness values & $-$ & $\checkmark$ \\
$\omega_i^\text{A}$ $\&$ $\omega_j^\text{B}$ & Gillespie movement rates for agents in group A and B & $-$ & $\checkmark$ \\
 & (Defined in the Appendices) & & \\
\hline
\end{tabular}
\label{tab:parameters}
\end{table*}

\subsection{Quantifying inequalities: the Gini coefficient} \label{sec:gini}

The Gini coefficient measures the degree of inequality in a given distribution and indicates how a particular distribution deviates from the uniform distribution. It is usually defined based on the \textit{Lorenz curve}, which shows the proportion of the total income of the population (represented in the vertical axis) that is cumulatively captured by the bottom x$\%$ of the population. The line at 45 degrees represents the perfect equality of a distribution. The Gini coefficient $G$ can be computed as \cite{amiel1999thinking}:
\begin{equation}
\begin{split}
G & =\frac{\sum_{i=1}^n \sum_{j=1}^n\left|x_i-x_j\right|}{2 \sum_{i=1}^n \sum_{j=1}^n x_j}=\frac{\sum_{i=1}^n \sum_{j=1}^n\left|x_i-x_j\right|}{2 n^2 \bar{x}},
\end{split}
\label{eq:Gini}
\end{equation}
where in our model $n$ is the number of individuals in the society, $x_i$ and $x_j$ are the fitness of the individuals $i$ and $j$, and $\bar{x}$ is the average of the distribution.

\begin{figure*}[btp!]
	\centering
        \begin{subfigure}[b]{0.45\textwidth}
		\includegraphics[width=\textwidth]{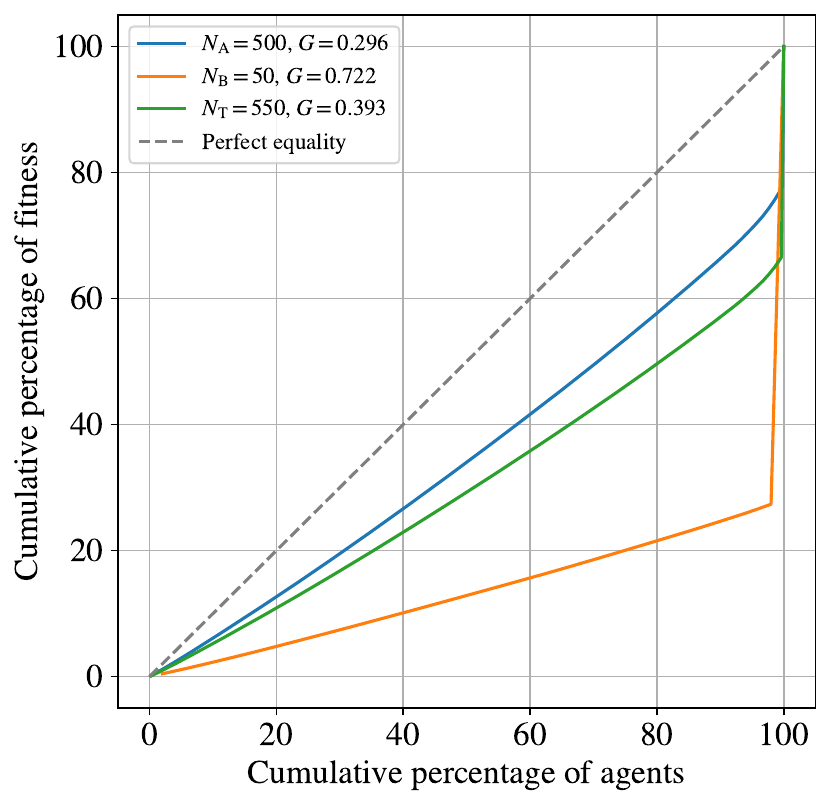}
            \caption{}
        \end{subfigure}
        \begin{subfigure}[b]{0.45\textwidth}
		\includegraphics[width=\textwidth]{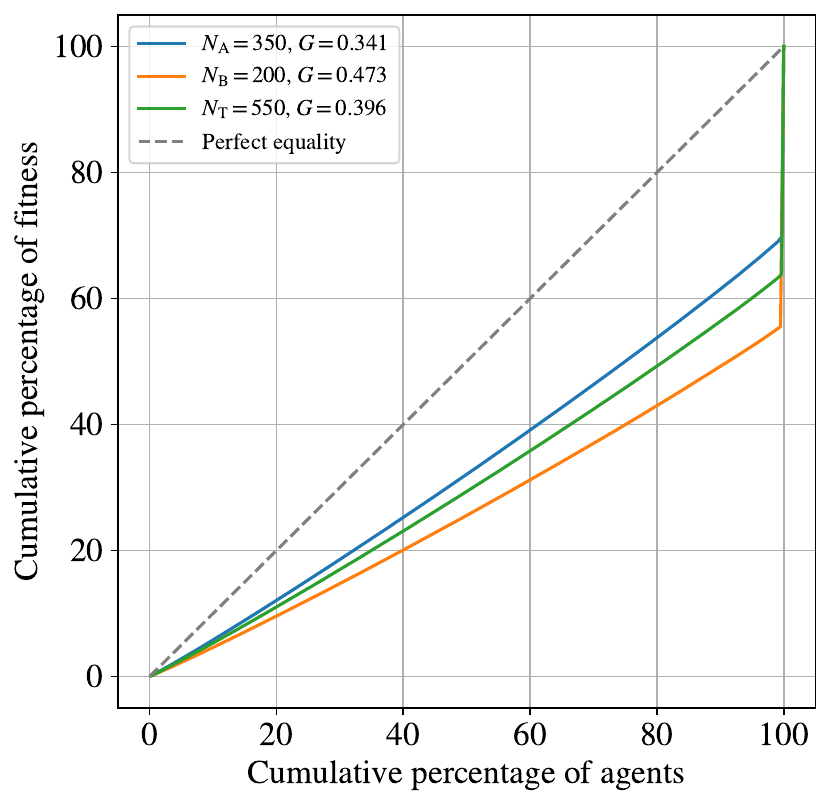}
            \caption{}
        \end{subfigure}
        \begin{subfigure}[b]{0.45\textwidth}
		\includegraphics[width=\textwidth]{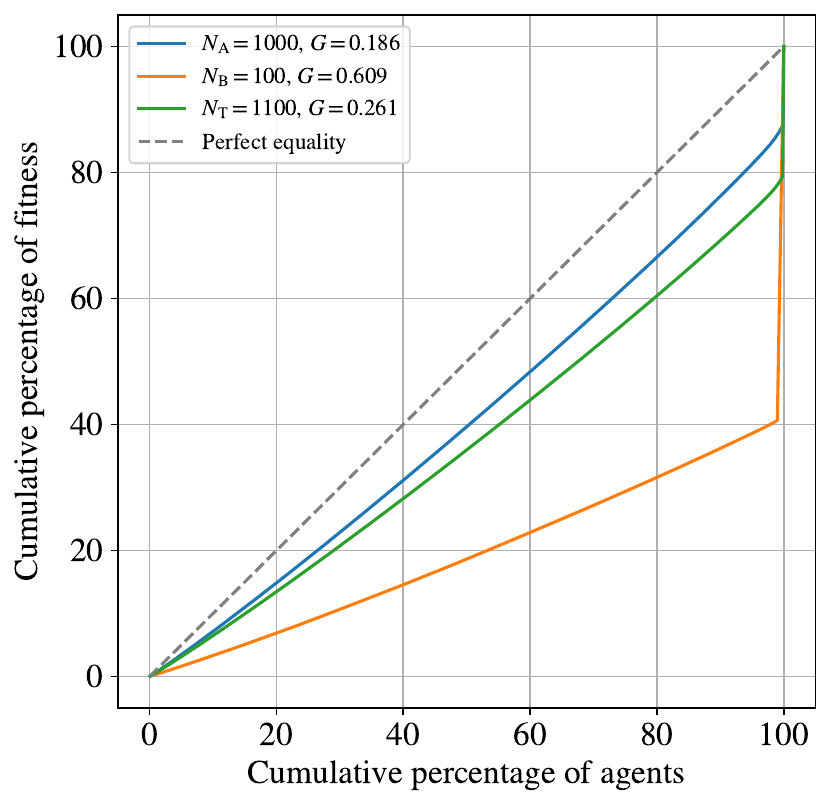}
            \caption{}
        \end{subfigure}
        \begin{subfigure}[b]{0.45\textwidth}
		\includegraphics[width=\textwidth]{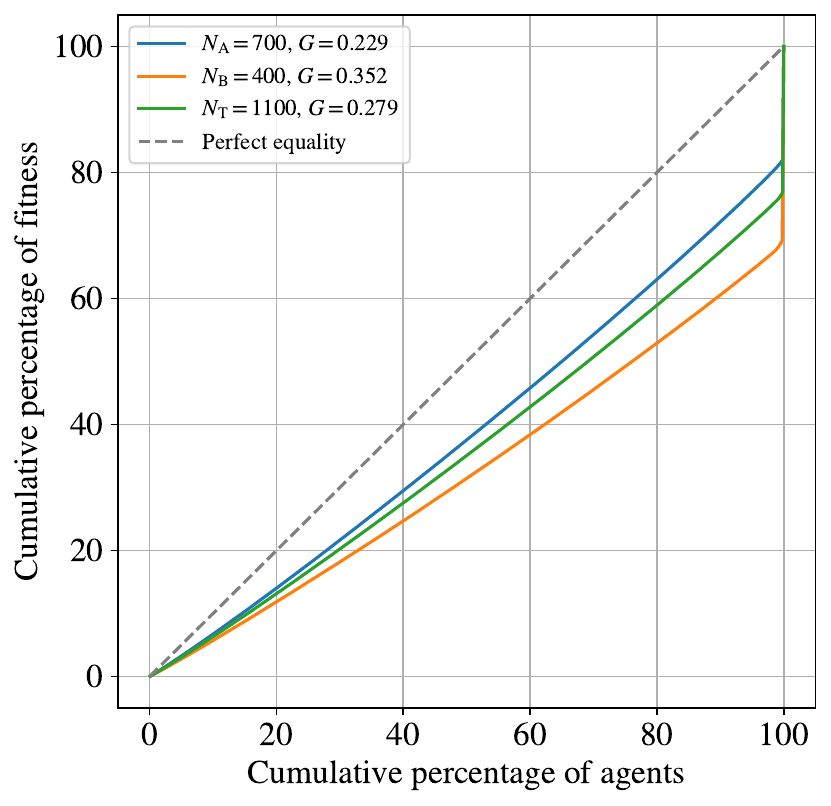}
            \caption{}
        \end{subfigure}
	\caption{\textbf{Lorenz curves and Gini coefficients for different number of agents.} Lorenz curves and Gini coefficients of one single trajectory at the stationary regime for $\eta= 5$ and $x=0.01$, randomly simulated on a $L\times L$ lattice ($L=25$). (a) $N_\text{A}=500$ and $N_\text{B}=50$. (b) $N_\text{A}=350$ and $N_\text{B}=200$. (c) $N_\text{A}=1\,000$ and $N_\text{B}=100$. (d) $N_\text{A}=700$ and $N_\text{B}=400$.}
	\label{f:GiniN}
\end{figure*}

Therefore, the Gini coefficient and the Lorenz curve for our simulations can serve to take a closer look at the inequalities as a function of the population for each group ($N_\text{A}$ and $N_\text{B}$). Figure \ref{f:GiniN} shows a fixed system size with the same combinations of $N_\text{T}$ and $N_\text{A}/N_\text{B}$ than in Figure \ref{f:N-L25-o}. 

It is evident that inequalities are highly sensitive to the number of agents that conform each group. Specifically, inequalities consistently appear more pronounced within the smaller group. Upon examining both groups depicted in Figure \ref{f:GiniN} (a) and Figure \ref{f:GiniN} (c), we observe that doubling $N_\text{T}$ while maintaining $N_\text{A}/N_\text{B}=10$ constant, leads to decreased inequalities within the minority group, the majority group and the overall society. Conversely, when holding the total society size $N_\text{T}=550$ constant, while the fraction $N_\text{A}/N_\text{B}=1.75$ is lowered, Figure \ref{f:GiniN} (b) shows that we can notably reduce inequalities within the minority group by slightly augmenting inequalities within the majority group. Interestingly, the inequalities within the entire system remain relatively unchanged in this scenario. A notable observation is that in all subfigures, there is a final large step attributable to the presence of a leader in each group. The height of this step provides information about the value of the maximum fitness. Particularly, the size represents the percentage of $F_\text{max}^\text{A}$ within its group, and the same for group B.

The Gini coefficient study reveals that inequality levels are significantly influenced by the population size of each group within a society. Precisely, inequalities are more pronounced in smaller groups. When the total population size is doubled while maintaining a fixed group ratio, inequalities decrease in both the minority and majority groups, as well as in the overall society. However, when the total population remains constant, but the group ratio is decreased making the groups more similar in size, inequalities within the minority group can be substantially reduced, even if this slightly increases inequalities within the majority group, without significantly affecting the overall societal inequality. This indicates that managing group sizes and ratios is a key factor in addressing and mitigating inequalities within a population.

\section{The role of $\eta$}\label{sec:roleofeeta}

%\section{Singular behavior with $\eta$}\label{sec:PT}

We now study $F_\text{max}^\text{A}$, ${F_\text{max}^\text{A}}/{F_\text{tot}^\text{A}}$ and $F_\text{tot}^\text{A}$ for the stationary regime and as a function of $\eta$. We repeat the analysis for several combinations of the number of agents $N_\text{A}$ and $N_\text{B}$, and also for the other group B. The results are sampled $250$ times during the stationary regime within a single simulation, each taken every $1\,000$ Gillespie time units. Data presented is also subject to an additional average across $50$ simulations.  To ensure normalized results, we set $\phi$ to 1 arbitrary unit for all simulations presented until the end.

\begin{figure}[btp!]
	\centering
        \begin{subfigure}[b]{0.41\textwidth}
		\includegraphics[width=\textwidth]{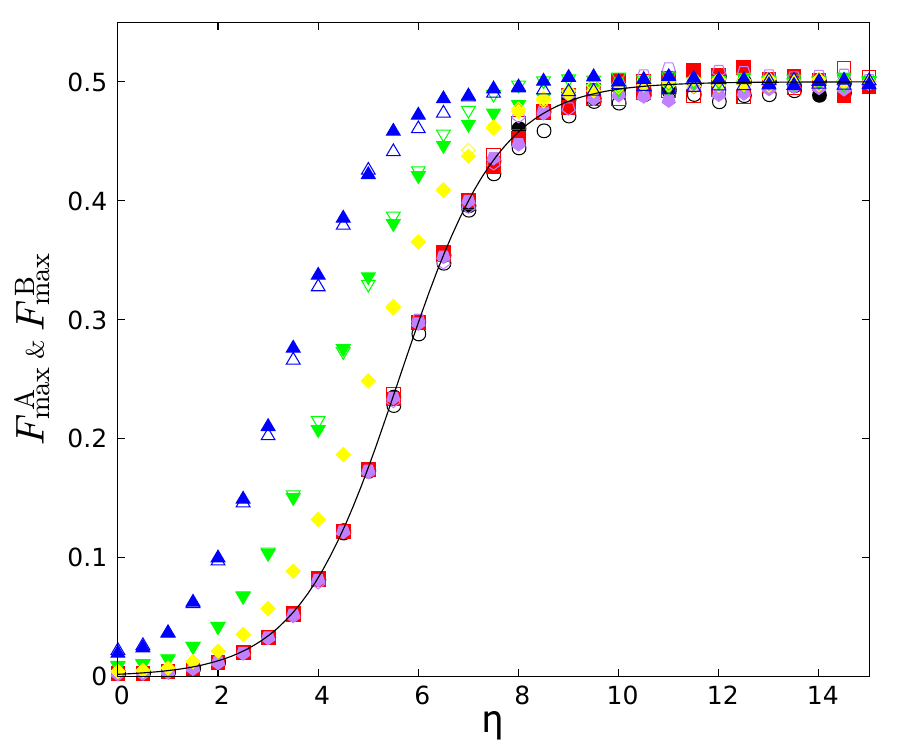}
            \caption{}
        \end{subfigure}
        \begin{subfigure}[b]{0.41\textwidth}
		\includegraphics[width=\textwidth]{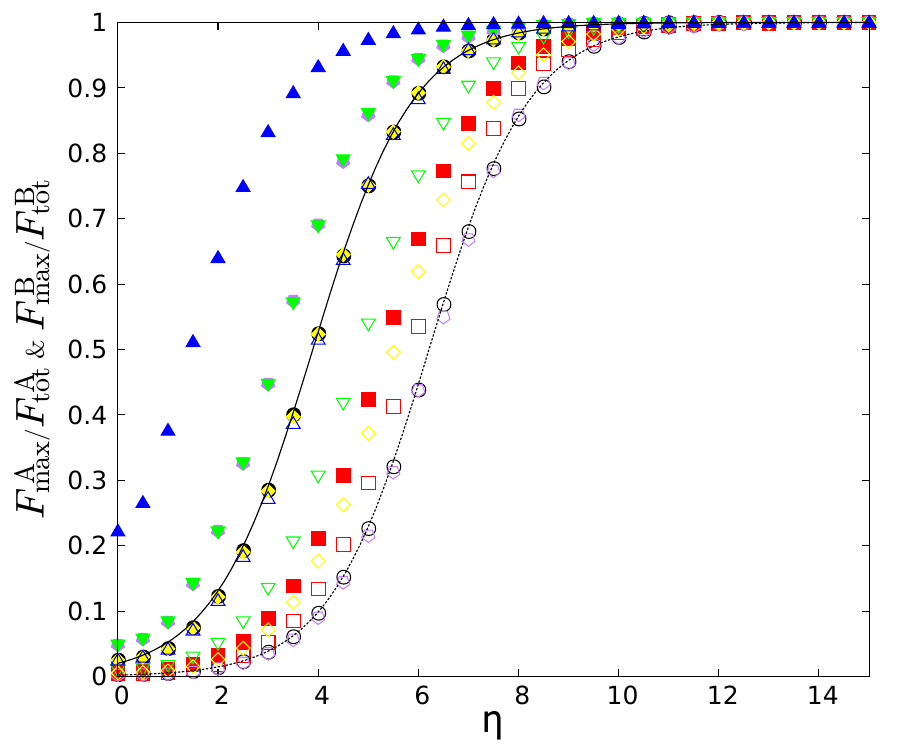}
            \caption{}
        \end{subfigure}
        \begin{subfigure}[b]{0.41\textwidth}
		\includegraphics[width=\textwidth]{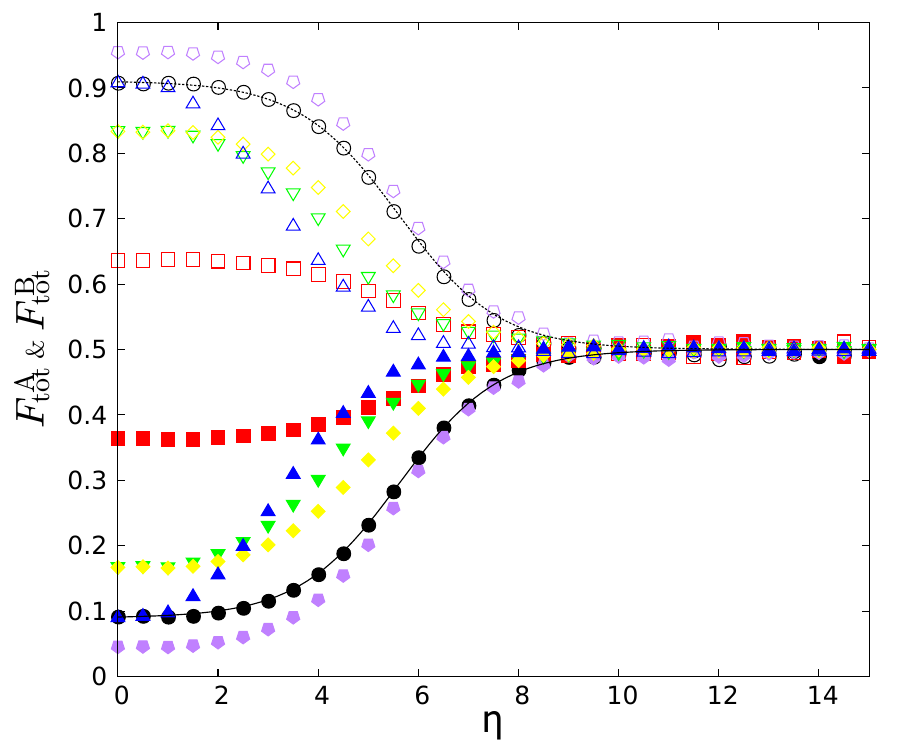}
            \caption{}
        \end{subfigure}
	\caption{\textbf{Total fitness and maximum fitness as a function of $\eta$.} The group A (empty symbols) and the group B (filled symbols) are in all cases plotted as a function of the parameter $\eta$, for $L=25$ and $x=0.01$: (a) Maximum agent fitness of each group. (b) Maximum agent fitness of each group normalized by the sum of all agent fitness of the same group. (c) Total fitness of each group. The simulations have been done for $N_\text{A}=500$ and $N_\text{B}=50$ (black circles), $N_\text{A}=350$ and $N_\text{B}=200$ (red squares), $N_\text{A}=525$ and $N_\text{B}=25$ (purple pentagons), $N_\text{A}=125$ and $N_\text{B}=25$ (green down triangles), $N_\text{A}=250$ and $N_\text{B}=50$ (yellow rhombuses), $N_\text{A}=50$ and $N_\text{B}=5$ (blue up triangles). Error bars are depreciable. Equations (\ref{eq:scalingfunctionfmaxftot}), (\ref{eq:scalingfunctionfmaxsumf}) and (\ref{eq:scalingfunctionftot}) are plotted for $N_\text{A}=500$ (dashed black lines) and $N_\text{B}=50$ (continuous black lines) to check the validity of the scaling laws.}
	\label{f:GlobalObservables}
\end{figure}

Figure \ref{f:GlobalObservables} (a) shows the maximum agent fitness of each group as a function of $\eta$. Essentially, these fitnesses represent the values of the leaders of each group. Figure \ref{f:GlobalObservables} (a) shows a possible phase transition; from an egalitarian society ($\eta=0$ all the agents have the same probability to win independent of their fitness, cf. Equation (\ref{eq:probbusiness})) to a hierarchical society, large $\eta$, one leader in each group is acquiring almost all the fitness, converging towards the common value of 1/2, regardless of the number of agents $N_\text{A}$ and $N_\text{B}$. For different combinations of $N_\text{A}$ and $N_\text{B}$, $F_\text{max}^\text{A}$ and $F_\text{max}^\text{B}$ evolve consistently with respect to $\eta$. This observation underscores that the leadership appearance remains unaffected by changes in the number of agents within each group. Nevertheless, curves shift towards large $\eta$ when the total number $N_\text{T}$ increases. Indeed, for societies composed by the same $N_\text{T}$, all curves overlap. This provides an initial insight into how these trends are influenced by $\eta$ and $N_\text{T}$. Finally, for sufficient large values of $\eta$, all considered examples of the observable collapse. We wonder if this is still true in the thermodynamic limit, that is, for $N_\text{T}\to \infty$, thus defining a critical value for $\eta$. Due to computational constraints, this study was unable to address this question.

Figure \ref{f:GlobalObservables} (b) represents the maximum agent fitness within each group, normalized by the total fitness of all agents within the same group. In essence, it signifies the leadership status within each group. Similar to the observation in Figure \ref{f:GlobalObservables} (a), a shift from an egalitarian to a hierarchical societal behaviour becomes apparent. Near $\eta=0$, this corresponds to the scenario with maximal equality, where all agents within their respective group possess the same fitness $1/N_\text{A}$, and likewise for the other group. On the opposite side, for large $\eta$, a clear leader emerges in each population. In this scenario, the curves shift towards the right as the number of agents $N_\text{A}$ (or $N_\text{B}$) increases. The curves are superimposed if the value of the number of agents is the same, regardless if corresponds to $N_\text{A}$ (the larger group) or $N_\text{B}$ (the smaller one). Once again, the question is whether the curves continue their rightward movement indefinitely or reach a point where they cease to shift at a certain $N_\text{A}$ and $N_\text{B}$.

Finally, Figure \ref{f:GlobalObservables} (c) shows the total fitness of each group as a function of the parameter $\eta$. The evolution of this observable exhibits some peculiarities. At small values of $\eta$, all individuals possess identical fitness, and consequently, this point is directly proportional to the number of agents in each group. Essentially, it is given by $N_\text{A}/{N_\text{T}}$, and the same for the other group. Hence, if the ratio $N_\text{A}/N_\text{B}$ is equal, simulations converge to the same value at $\eta=0$. In contrast to previous observations, the behaviour of the curves is diverse depending on the number of agents. Also, looking at larger $\eta$, it seems to exist a singular $\eta$ around 8 that increases with $N_\text{T}$, where the fitness distribution between groups does not depend on the number of individuals. In this case, each group shares an equal amount of fitness.

All in all, results provided in Figure \ref{f:GlobalObservables} suggests that from a particular $\eta$ value onwards, a significant portion of the total fitness is primarily kept by the leader individuals. As we mention above, we do not have enough computational power to find the thermodynamic limit.
%\textit{Appendix \ref{sec:observablesagents}} adds additional information of how the first two observables evolve concerning the number of agents across various values of $\eta$. 
The question that arises at this point is what analytical functional dependence have all these observables in relation to $\eta$ and the number of agents.

\section{Scaling laws} \label{sec:USL}

We first investigate how the maximum agent fitness and the maximum agent fitness normalized by the total fitness of all agents within the same group, can be represented by the same curve, regardless of the number of agents. Both quantities have the same sigmoidal shape as a function of $\eta$. We thus suggest a sigmoidal fit of $F_\text{max}^\text{A}$ for several values of $N_\text{T}$ (see Figure \ref{f:GlobalObservables} (a)):
\begin{equation}
    g(\eta,N_\text{T})=\frac{1}{2+e^{-a_0^g(N_\text{T})(\eta-\eta_0^g(N_\text{T}))}}.
    \label{eq:Fitfmaxftot}
\end{equation}
We also suggest another (very similar) sigmoid fit of ${F_\text{max}^\text{A}}/{F_\text{tot}^\text{A}}$ for several values of $N_\text{A}$ (see Figure \ref{f:GlobalObservables} (b)): 
\begin{equation}
    h(\eta,N_\text{A})=\frac{1}{1+e^{-a_0^h(N_\text{A})(\eta-\eta_0^h(N_\text{A}))}}.
    \label{eq:Fitfmaxsumf}
\end{equation}
The same expressions would apply for the group B. We can then estimate $a_0^g$ and $\eta_0^g$ as a function of $N_\text{T}$, and $a_0^h$ and $\eta_0^h$ as a function of $N_i$, to see their dependence on the number of agents. The fitted parameters behave as follows:
\begin{equation}
\begin{aligned}
& a_0^g \approx 1 \; \forall N_\text{T} \\
& \eta_0^g (N_\text{T}) = \ln(N_\text{T}-2) \\
& a_0^h \approx 1 \; \forall N_\text{A} \\
& \eta_0^h (N_\text{A}) = \ln(N_\text{A}-1). 
\end{aligned}
\label{eq:fittedparamters}
\end{equation}
And the same relationships for the group B. The reason for these $-2$ and $-1$ values inside the logarithms are in order to satisfy the boundary condition for $\eta=0$, where the fitted functions have to take the values of $1/N_\text{T}$ and $1/N_\text{A}$ (or $1/N_\text{B}$), the so-called the egalitarian society.

\begin{figure}[btp!]
	\centering
        \begin{subfigure}[b]{0.45\textwidth}
		\includegraphics[width=\textwidth]{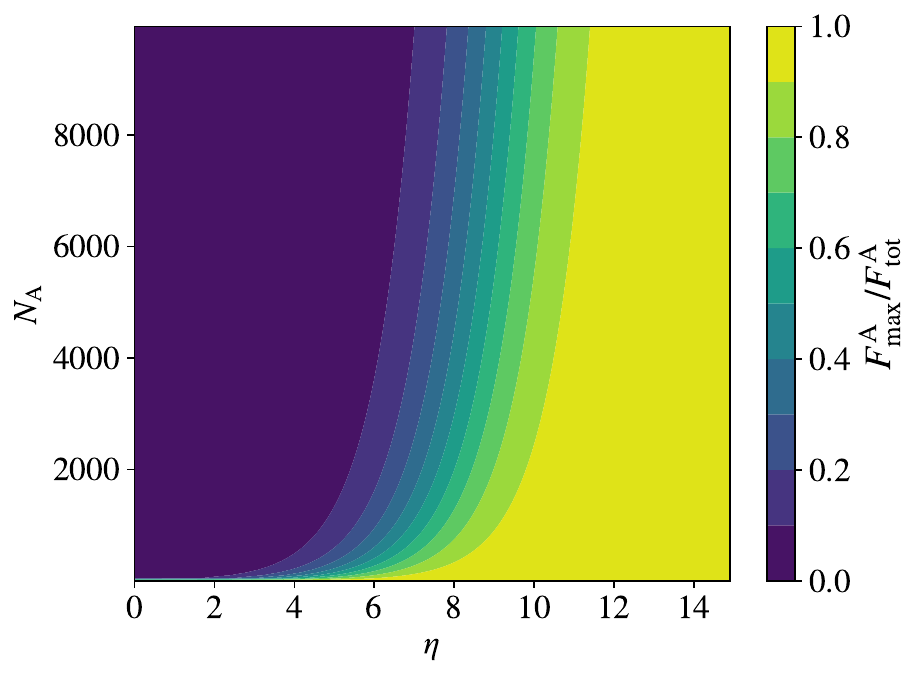}
            \caption{}
        \end{subfigure}
        \begin{subfigure}[b]{0.45\textwidth}
		\includegraphics[width=\textwidth]{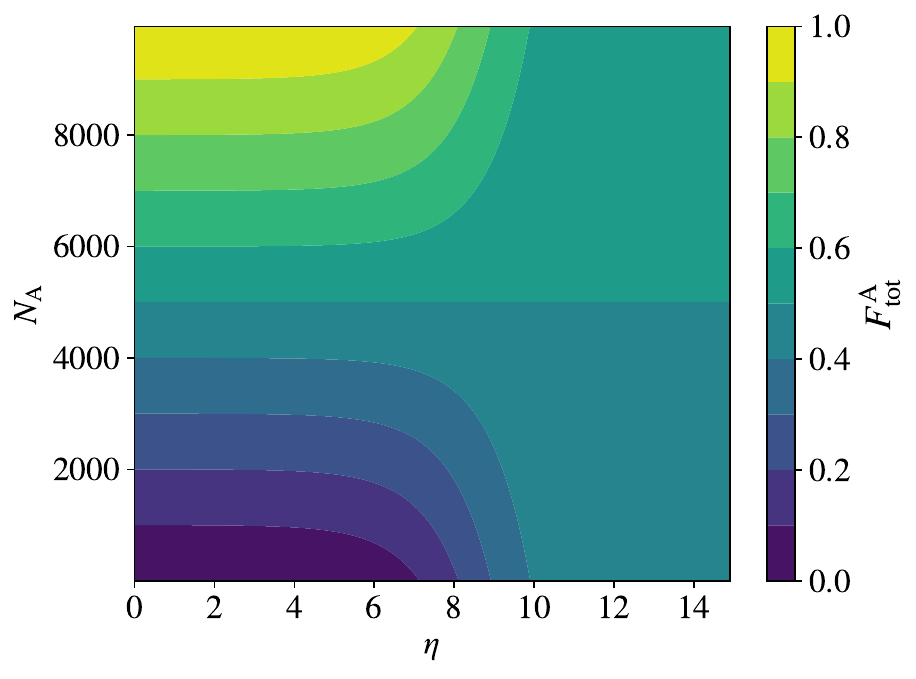}
            \caption{}
        \end{subfigure}        
	\caption{\textbf{Heat maps of the scaling functions for the maximum fitness normalized by the sum agent fitness and the total fitness of each group.} (a) 2D plot of the scaling function given by Equation (\ref{eq:scalingfunctionfmaxsumf}) in colour, as a function of the parameters $\eta$ and the number of agents of the group A. (b) 2D plot of the scaling function given by Equation (\ref{eq:scalingfunctionftot}) in colour, as a function of the parameters $\eta$, the number of agents of the group A and a fixed total number of agents $N_\text{T}=10\,000$.
    }
	\label{f:scaling}
\end{figure}

Functions $F_\text{max}^\text{A}$ and ${F_\text{max}^\text{A}}/{F_\text{tot}^\text{A}}$ can be therefore effectively described by sigmoidal functions depending on the free parameter $\eta$, the total number of agents $N_\text{T}$, and the number of agents in each group as:
\begin{eqnarray}
    F_\text{max}^\text{A}(\eta,N_\text{T})=\frac{1}{2+(N_\text{T}-2)e^{-\eta}} ,
    \label{eq:scalingfunctionfmaxftot}\\
    \frac{F_\text{max}^\text{A}}{F_\text{tot}^\text{A}}(\eta,N_\text{A})=\frac{1}{1+(N_\text{A}-1)e^{-\eta}} ,
    \label{eq:scalingfunctionfmaxsumf}
\end{eqnarray}
and the same expressions apply for the group B. These sigmoidal shapes can be related to a phase transition from egalitarian to hierarchical societies for each group as a function of the control parameter $\eta$. Figure \ref{f:scaling} (a) shows a phase diagram of the Equation (\ref{eq:scalingfunctionfmaxsumf}) to see this clear behavioural change depending on these both parameters. In the blue region, the absence of a leader is observed, whereas transitioning to the yellow region signifies the emergence of either multiple leaders or a predominant leader, ultimately culminating in the establishment of a singular leader within the yellow region.

Furthermore, the total fitness of each group, depicted in Figure \ref{f:GlobalObservables} (c), can be readily determined as the ratio between Equations (\ref{eq:scalingfunctionfmaxftot}) and (\ref{eq:scalingfunctionfmaxsumf}). Accordingly, the expression for $F_\text{tot}^\text{A}$ is derived as a function of the free parameter $\eta$, the total number of agents $N_\text{T}$, and the number of agents in each group as:
\begin{equation}
    F_\text{tot}^\text{A}(\eta,N_\text{T},N_\text{A})=\frac{F_\text{max}^\text{A}(\eta,N_\text{T})}{F_\text{max}^\text{A}/F_\text{tot}^\text{A}(\eta,N_\text{A})}=\frac{1+(N_\text{A}-1)e^{-\eta}}{2+(N_\text{T}-2)e^{-\eta}}=\frac{e^{\eta}+N_\text{A}-1}{2e^{\eta}+N_\text{T}-2},
    \label{eq:scalingfunctionftot}
\end{equation}
and the expression apply for the group B. This function highlights the behavioral transition of the total fitness in each group as a function of the control parameter $\eta$, while satisfying the boundary conditions at $\eta = 0$ and for large values of $\eta$. Figure \ref{f:scaling} (b) shows a phase diagram of the Equation (\ref{eq:scalingfunctionftot}) to see this clear behavioural change depending on these three parameters. In the blue and yellow regions, the entire total fitness $\phi$ belongs to the group A or B, whereas transitioning to the green regions signifies that the overall fitness of the entire society is distributed among both groups,  ultimately culminating in the establishment of a equal distribution within the green region. This transition can emerge when the number of agents of the groups tend to the same size and also for large enough values of $\eta$, where group sizes becomes irrelevant.

\section{Discussion}\label{sec:discussion}

This work builds upon the Bonabeau model. The original Bonabeau model examines the formation of hierarchies in societies through interactions among identical agents \cite{bonabeau_phase_1995}. Bonabeau agents gain or lose fitness based on the outcome of pairwise interactions. The interest in the Bonabeau model is in providing mechanisms and certain conditions that enables the emergence of social hierarchies in a society. The original model combines a relaxation process (individual fitness decays towards zero over time) and a competition process (fitness differences between agents determine the probability of winning). The balance, or imbalance, among these two mechanisms determines the presence of social hierarchies, and the system undergoes a phase transition. When the density of agents $\rho$ (determining the frequency of interactions) is high enough, the competition term dominates, allowing the fitness differences between identical agents to increase, and thus the emergence of hierarchies. Conversely, if the density $\rho$ is too low, interactions are infrequent, and the relaxation term dominates, resulting in an egalitarian society with near zero fitness values. This transition is often characterized by parameter $\sigma$, which measures the variance in the number of wins among agents in respect the egalitarian case. As density increases, $\sigma$ increases, signaling the formation of a hierarchical structure and the emergence of classes \cite{tsujiguchi_self-organizing_2007, odagaki_self-organizing_2006}.

Our work removes the Bonabeau relaxation component but adds a second group of agents. Our model only allows interactions with agents from the opposite group. If one focuses the attention to one group, the addition of a second group is comparable to a quite peculiar relaxation mechanism. As a result, our framework can also describe the emergence of social hierarchies, while taking a totally different perspective. It then becomes hard to compare our model with the Bonabeau model in a trivial manner. In our model, the transition from an egalitarian to a hierarchical society relies on the inverse of the temperature $\eta$ which grades the fitness difference between interacting agents. Larger $\eta$ thus favors probability to win of agent with largest normalized fitness values (cf. Eq. (\ref{eq:probbusiness})). Also, because our model does not include a relaxation term, the density of agents $\rho$ does not play a key role to the dynamics. It then only regulates the speed of the process (as shown in Figure \ref{f:fmax-s}). Indeed, social hierarchies in the Bonabeau emerge for non-zero $\eta$. Notably, the asymptotic limits of $\eta$ in our model correspond directly to $\rho$ limits in the Bonabeau model: when $\eta$ approaches to zero, it mirrors $\rho$ approaching also zero, yielding an egalitarian society. Conversely, as $\eta$ increases in our model, the system transitions to a hierarchical state, as happens for $\rho$.

Another relevant aspect to confront with the original Bonabeau model is related to the fitness definition. The Bonabeau model takes absolute fitness values of the interacting agents to compute the probability of winning. In contrast, our model considers normalized fitness values. This is motivated by the fact that we were wanting to reflect the relative prestige of agents within their respective groups. If we were to rely on absolute fitness values instead of normalized ones, fitness of one group would be entirely transferred to the other group. This would result in one group being reduced to zero total fitness, while the other group would stabilize with a distribution, with no further fitness exchange. Normalized fitness values introduce the concept of prestige within each group. For $\eta$ approaching zero, the system becomes insensitive to prestige, and interactions are egalitarian. Conversely, as $\eta$ increases, the probability of winning is entirely determined by prestige within the group. The agent possessing the highest prestige is consistently winning. Furthermore, being the fitness exchange proportional to the opponent's fitness ensures that the total fitness $\phi$ within the system remains constant over time. This is another notable departure from the Bonabeau model, where the magnitude of individual fitness can grow indefinitely. Finally, if the fitness exchange were fixed as it is generally taken in the Bonabeau rather than proportional to the opponent's fitness, the conservation property would also be lost. The system would then exhibit dynamics more akin to the Bonabeau model, requiring additional constraints if we want to maintain stability in the system. In such a scenario, the dynamics would produce negative fitness values.

Our model also does take different aggregate observables than the those generally used in the Bonabeau model. The Bonabeau model considers the variance of the number of wins, $\sigma$. For our model, it is more interesting and meaningful to consider the fitness of individual agents. This approach provides a more granular understanding of how hierarchical structures emerge and evolve at the agent level. The use of the Gini index also reveals different classes within each group in terms of the cumulative percentage of agents, as reflected in the varying slopes of the Lorenz curves for different regimes (see Fig. \ref{f:GiniN}). In this way, the internal stratification of classes within each group can be better observed and compared. Also, the curves also reveal the emergence of a dominant leader within each group in its extreme value. The relative fitness of these leaders is strongly influenced by the size of their respective groups, with smaller groups producing leaders of significantly higher prestige. These are particular features which can only retrieved by considered fitness aggregate values and more sophisticated metrics such as the Gini index.

The dynamics of the model can also be interpreted through the lens of a bipartite graph. In this interpretation, agents in group A and group B form two distinct sets of nodes, and interactions occur only between nodes from opposite sets. The random interactions in the square lattice representation of the model can be mapped onto a fully connected bipartite graph framework, where pairs of nodes are selected randomly to interact \cite{malarz_bonabeau_2006}. This perspective highlights that the core interaction rules of the model are independent of the spatial constraints imposed by the lattice. However, the spatial dynamics introduce additional factors, such as the density of agents and the probability of encounters, which influence the time scales of the system's evolution \cite{fujie2011self}. For instance, lower densities reduce the probability of interactions, slowing down the convergence to the stationary state. In contrast, higher densities accelerate this process. The bipartite graph representation thus replaces spatial factors such as the agent density of the lattice-based model by a parameter that governs the interaction probability. Viewing the model as a bipartite graph problem offers additional insights into the role of interaction probabilities and time scales, while maintaining the same stationary-state results observed in the lattice-based version.

Structuring interactions between two distinct groups can also drive the transition from an egalitarian society to a hierarchical one. This approach that prevents interactions within the same group can also amplify fitness differences and foster the concentration of fitness among few agents. By focusing on fitness conservation, group-based prestige interactions, and proportional fitness exchange, our framework provides novel insights into the mechanisms underlying inequality and dominance in structured populations.

\section{Concluding remarks}\label{sec:conclusions}

The formation of hierarchies in societies is an intriguing topic in which agent-based modelling approach can provide important insights. Hierarchies are assumed to be built in a bottom-up approach and based on interactions between agents. Hierarchies are widely present in human societies, and they appear in contexts such as cities where interactions intensively take place \cite{moro_mobility_2021}. The formation of hierarchies in urban areas has serious consequences in terms of segregation and inequality \cite{checa_residential_2021, espin-noboa_inequality_2022}. Cities are organized in systematic manners \cite{bettencourt_2013,bettencourt_2021} and they shape human interactions \cite{oliveira_2017, lobo_2020}. In fact, face-to-face interaction is a basic human behaviour, modelling how people form and keep social groups by segregating themselves from others \cite{goffman_1967,kendon_1975,duncan_1977,bargiela_2009}. The lack of social cohesion can undermine the social fabric of cities and dramatically affect the economic, social and health conditions of people living in urban areas \cite{florida_2017, finneran_2003}. Macroscopic social hierarchies can spontaneously emerge from micro-motives in human daily habits. Measurable attributes such as income, education, occupation, and wealth are used as indicators of socioeconomic status, either individually or in combination to create composite indicators of socioeconomic standing \cite{vickie_2007}. However, our knowledge of the interplay between group dynamics and the emergence of inequalities in social systems is still limited and quantitatively researched \cite{oliveira_2022}. Although there are many studies on the tendency of human beings to form groups that are socially cohesive, the lack of large-scale data has left many questions about social groups undetermined. For instance, in \cite{zignani_2019} the study of the nature of urban groups and a methodology for their identification is developed thanks to anonymized mobile phone datasets, such as Call Detail Records.

A better understanding of the formation of hierarchies and their consequences is what has motivated us to extend the Bonabeau model by introducing a second group in the society. The model only allows for interactions among agents that belong to different groups. Under certain conditions, Monte Carlo simulations show that for a broad range of values of the model, the fitness of the agents of each group present a decay in time except for one or few agents which capture almost all the fitness of the system. A clear behavioural change in several ways of representing the fitness of the system is observed when the control parameter $\eta$ is changed. The phenomena can be understood as a phase transition from an egalitarian to a hierarchical society. The results remain robust with respect to the system size and depend solely on the number of agents within each group and their ratio. In contrast to the original Bonabeau model \cite{bonabeau_phase_1995}, our model shows that the degree of the inequalities do not depend on the overall density of individuals in the society or within each group. Instead, it is determined by the number of individuals and it is then basically a question of number of interactions. Scaling functions for the maximum agent fitness, the maximum agent fitness normalized by the sum of all agent fitness within the same group, as well as for the total fitness of each group are found and the mathematical expressions obtained can be useful to extend the results to larger systems.

The model may be related to different real-world situations and it is thus possible to further reflect on the outcomes under certain circumstances. The lattice might be a proxy of a city in which inhabitants are constantly moving around. One can imagine that the two groups represent people and businesses and the only interactions possible are transactions among people and businesses. Then we can look at different scenarios for businesses. For small $\eta$, all businesses share the same fitness, which this situation is known as \textit{perfect competition}. Increasing $\eta$, a dominant business emerges, but not with all the fitness of its group. We are in the \textit{monopolistic competition} regime and an \textit{oligopoly} gives rise. Finally, when $\eta$ is large enough, all fitness is concentrated into to a single business (a \textit{monopoly}). The same reasoning might apply with two competing criminal gangs sharing the same physical space (eventually a city). The model might thus be telling us that the more fights the more important would be the role of each gang leader while other members of each gang would be those losing all fitness and all power or wealth. Another situation which is becoming more and more frequent is the competition for urban space between tourists and neighbors and just by the existence of these interactions many side-effects are just happening such as gentrification of the center of the cities. In this case, the model can help to further reflect how only few of each of group (tourists and neighbours) hold most of the privileges. Other situations can be also imagined when two very different social groups are in interaction. To compare numerical simulations of the model with real data is expected to be a future work, by for instance taking household income as a proxy for societal status in a city \cite{hickey_self-organization_2019,hong_measuring_2021}.

%In the present day, disparities in living conditions within regional areas are on a constant upward trajectory, with a noticeable spatial manifestation through residential segregation. Thanks to the refinement of technology, today many real social indicators about the population can be archived, such as the level of education, risk of poverty, household incomes, and so on. All these indicators can be analysed from various perspectives to explore this argument and classify every indicator’s explanatory potential \cite{checa_residential_2021}. The starting point for investigation is the temporal evolution of each agent's fitness as the model parameters are altered.

The model can be extended in many ways such as other winners' probabilities \cite{hickey_self-organization_2019,ispolatov_wealth_1998}, the redistribution of the total fitness applying a relaxation term in the temporal evolution rules, and so on. It is indeed possible to add other sort of interactions, to allow other movements (not only a random walk) or to add constraints  \cite{tsujiguchi_self-organizing_2007, odagaki_self-organizing_2006, posfai_talent_2018,woolcock_fitness_2017}, to use time-varying networks \cite{kawakatsu_emergence_2021} and even to increment the number of groups. Also, game-theoretical models capture very flexible situations where cooperation among selfish agents can emerge \cite{lee_cooperation_2011, lozano_cooperation_2020} and analytical studies could be implemented \cite{lacasa_bonabeau_2006}.

\printcredits

\section*{Competing interests}
The authors declare no competing interests.
\section*{Acknowledgments}
This work has been partially funded by MCIN/AEI/ 10.13039/501100011033, grant number PID2019-106811GB-C33 (JP and MM); by MCIN/AEI/ 10.13039/501100011033 and by ``ESF Investing in your future'', grant number PRE2020-093266 (MS); by MCIN/AEI/ 10.13039/501100011033 and by ``ERDF A way of making Europe'', grant number PID2022-140757NB-I00. We also acknowledge the support of Generalitat de Catalunya, grant number 2021SGR00856.
\section*{Appendices}
\appendix

\section{Fitness normalization} \label{sec:Norm}
\begin{figure}[btp!]
	\centering
        \begin{subfigure}[b]{0.45\textwidth}
		\includegraphics[width=\textwidth]{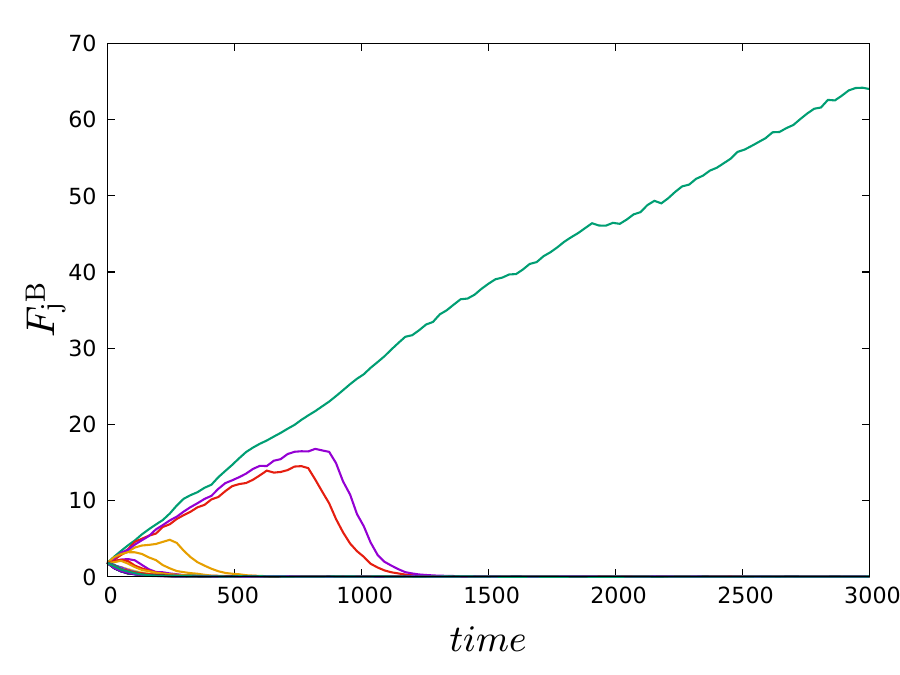}
              \caption{}
        \end{subfigure}
        \begin{subfigure}[b]{0.45\textwidth}
		\includegraphics[width=\textwidth]{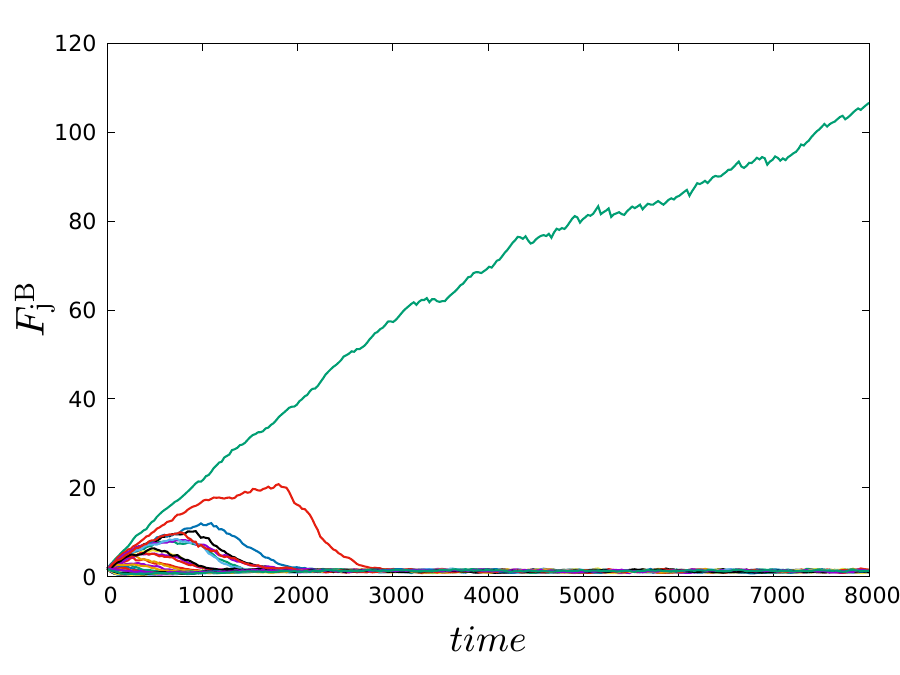}
              \caption{}
        \end{subfigure}
	\caption{\textbf{Temporal evolution of the fitnesses of group B for two different normalizations.} Time evolution of all $F_{j}^\text{B}$ for $N_\text{A}=500$ and $N_\text{B}=50$ individuals, $\eta= 5$ and $x=0.01$, randomly simulated on a $L\times L$ lattice ($L=25$) and two different normalizations. (a) The first $3\,000$ time steps using the linear scaling to a range of $[-1,1]$. (b) The first $8\,000$ time steps using the mean normalization.}
	\label{f:5-001-L25-norm}
\end{figure}

To compare the prestige/reputation between two agents of opposite groups when a random interaction happens, we normalize them under their respectively sample. Therefore, when an exchange is possible, performing this normalization we will not compare directly the fitness but their prestige which each has into its particular group. We have proved the following scenarios:
\begin{enumerate}
    \item \textbf{Scaling (min-max normalization):}
    \begin{equation}
        \hat{F}_{j}^\text{B}(t)=\frac{{F}_{j}^\text{B}(t)-{F}_\text{min}^\text{B}(t)}{{F}_\text{max}^\text{B}(t)-{F}_\text{min}^\text{B}(t)} \;,\;[0,1]
        \label{eq:normalization1a}
    \end{equation}
    \item \textbf{Linear scaling to a range:}
    \begin{equation}
        \hat{F}_{j}^\text{B}(t)=a+\frac{({F}_{j}^\text{B}(t)-{F}_\text{min}^\text{B}(t))(b-a)}{{F}_\text{max}^\text{B}(t)-{F}_\text{min}^\text{B}(t)} \;,\;[a,b]
        \label{eq:normalization2a}
    \end{equation}
    \item \textbf{Mean normalization:}
    \begin{equation}
        \hat{F}_{j}^\text{B}(t)=\frac{{F}_{j}^\text{B}(t)-\langle F_j^\text{B}(t) \rangle}{{F}_\text{max}^\text{B}(t)-{F}_\text{min}^\text{B}(t)}
        \label{eq:normalization3a}
    \end{equation}
\end{enumerate}
where $a$ and $b$ are free integer numbers and $\langle F_j^\text{B}(t) \rangle$ is the mean of the fitnesses vector. And the same equations for $\hat{F}_{i}^\text{A}(t)$. In Figure \ref{f:5-001-L25} (a) and Figure \ref{f:5-001-L25-norm}, we show that applying these three types of normalizations, the results are qualitatively the same. For simplicity, we have chosen the first one, Equation (\ref{eq:normalization1a}).

\section{Monte Carlo methods}\label{sec:MC}
In this Appendix, a brief review of the Monte Carlo setup is explained:
\begin{enumerate}
    \item The lattice is equipped by Periodic Boundary Conditions (PBC). In terms of a city, this could be understood as a simplification of that people travelling from in to out of the city, and vice versa.
    \item A residence time algorithm, also called Gillespie algorithm, is applied to reproduce the time when a jump happens \cite{raul}. The movement rate of each agent of the group A, $\omega_i^\text{A}$, is settled to $1$, and the same for the group B, $\omega_j^\text{B}=1$. Therefore, the total rate when a given agent of the first group moves is $\Omega_\text{A}=N_\text{A}\omega_i^\text{A}=N_\text{A}$ and the same for the second group $\Omega_\text{B}=N_\text{B}\omega_j^\text{B}=N_\text{B}$ . Then, the global rate $\Omega_\text{T}$ when any move happens corresponds to the sum of both: $\Omega_\text{T}=\Omega_\text{A}+\Omega_\text{B}$. The time $t$ in our simulations will be in units of $\omega$, and the time step to the next movement is randomly simulated as:
    \begin{equation}
        \Delta t=-\frac{\ln u}{\Omega_{\text{T}}},
        \label{eq:deltatime}
    \end{equation}
    where $u\sim U(0,1)$ is a uniform random number between 0 and 1. After given that a movement has occurred, we randomly identify which group of agent performs it according to the total rates of group A and B. If the movement rates of all agents of one group would settle to zero, the system could be understood as a business-client interaction, where one of the two groups is static.
    \item We register one point every $N_\text{movements}$ or $dt$ to reduce storage needs.
    \item We have observed that changing the initial distribution of fitness among the agents, the tendency towards the stationary regime does not change.
\end{enumerate}

\section{System size dependence} \label{sec:Systemsize}
\begin{figure}[btp!]
	\centering
        \begin{subfigure}[b]{0.45\textwidth}
		\includegraphics[width=\textwidth]{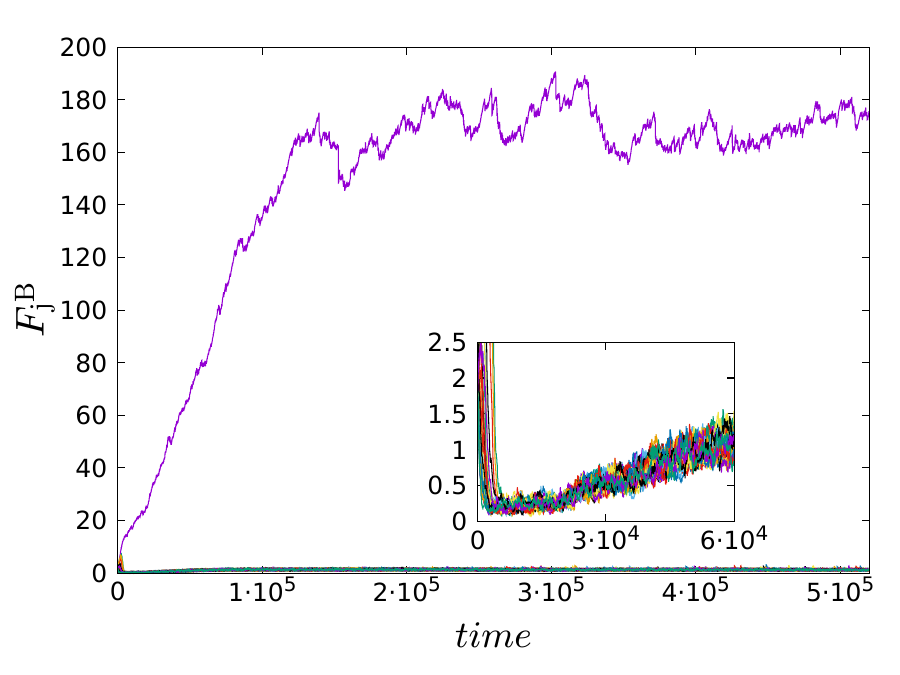}
              \caption{}
        \end{subfigure}
        \begin{subfigure}[b]{0.45\textwidth}
		\includegraphics[width=\textwidth]{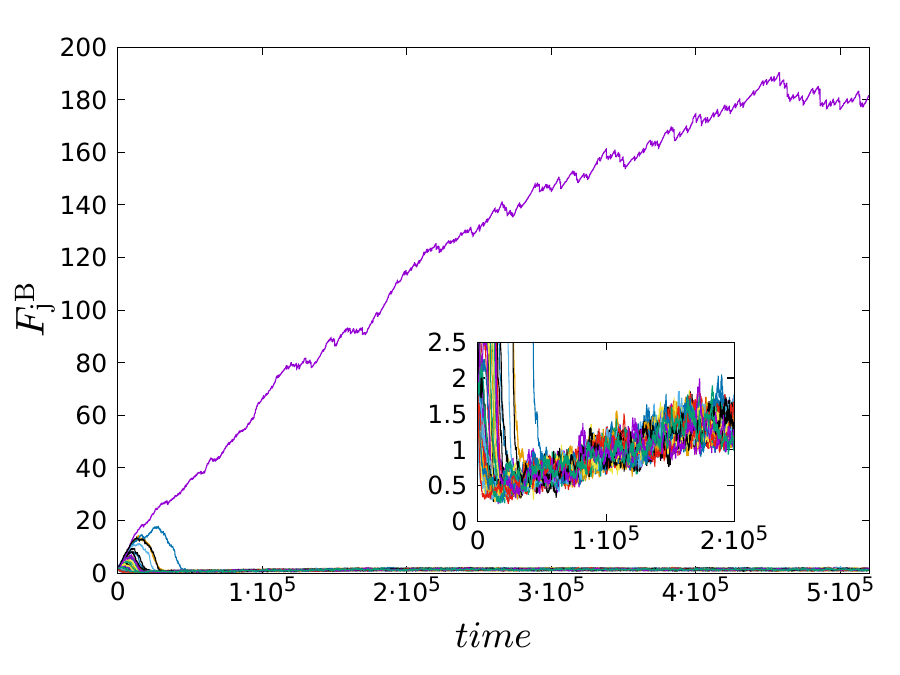}
              \caption{}
        \end{subfigure}
		\caption{\textbf{Temporal evolution of the fitnesses of group B for two systems sizes, $\eta=5$ and $x=0.01$.} Time evolution of all $F_{j}^\text{B}$ for $N_\text{A}=500$ and $N_\text{B}=50$ individuals, $\eta= 5$ and $x=0.01$, randomly simulated on two 2D lattice sizes. (a) $L=65$, (b) $L=120$.}
	\label{f:5-001-L-s}
\end{figure}

We here illustrate the connection between the stationary regime and the system size, as well as its dependence on the time required to reach this state. First, trajectories for the fitness of the group B are plotted for two distinct values of the system size $L$ in Figure \ref{f:5-001-L-s}. We see how the leader appears lately (the second case) as well as the time required to reach the stationary regime.

\begin{figure}[btp!]
	\centering
		\includegraphics[width=0.48\textwidth]{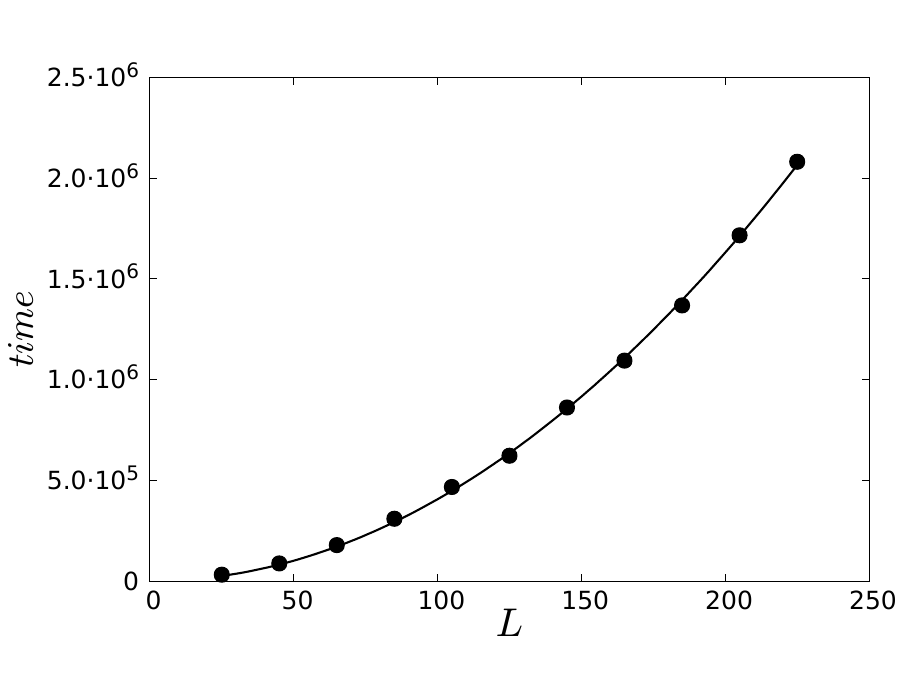}
		\caption{\textbf{Scaling time when a dominant agent of the group B reach the stationary regime as a function of the system size.} Scaling time when the maximum fitness value of the group B reaches the average stationary value as a function of the system size $L$, for $N_\text{A}=500$ and $N_\text{B}=50$ individuals, $\eta= 5$ and $x=0.01$. Numerical values averaged under 100 different runs (filled circles) and a potential fit of the form: $t\sim aL^b$ (solid line) is plotted. Error bars are depreciable.}
	\label{f:timescaling}
\end{figure}

Besides, in Figure \ref{f:timescaling} we compute the stationary numerical times for many system sizes. We have obtained these times by averaging over $100$ different runs when the maximum value in the group reaches the average maximum fitness computed in this stationary regime. For the parameters showed in Figure \ref{f:timescaling}, the average value of the maximum fitness of the group B computed in the stationary regime for all $L$ is around 175 (see Figure \ref{f:fmax-s}). The obtained values for these particular parameters of the model from the least squares fit $t\sim aL^b$  have been: $a=39.5\pm4.0$ and $b=2.0\pm0.1$. Therefore, this scaling behaviour could be interpreted as a diffusion process of the agents' fitness in the space, where larger is the territory larger is the time to have an encounter, and consequently, larger the time of leader emergence.

\section{Exchange factor dependence} \label{sec:exchangefactor}
\begin{figure}[btp!]
	\centering
        \begin{subfigure}[b]{0.45\textwidth}
		\includegraphics[width=\textwidth]{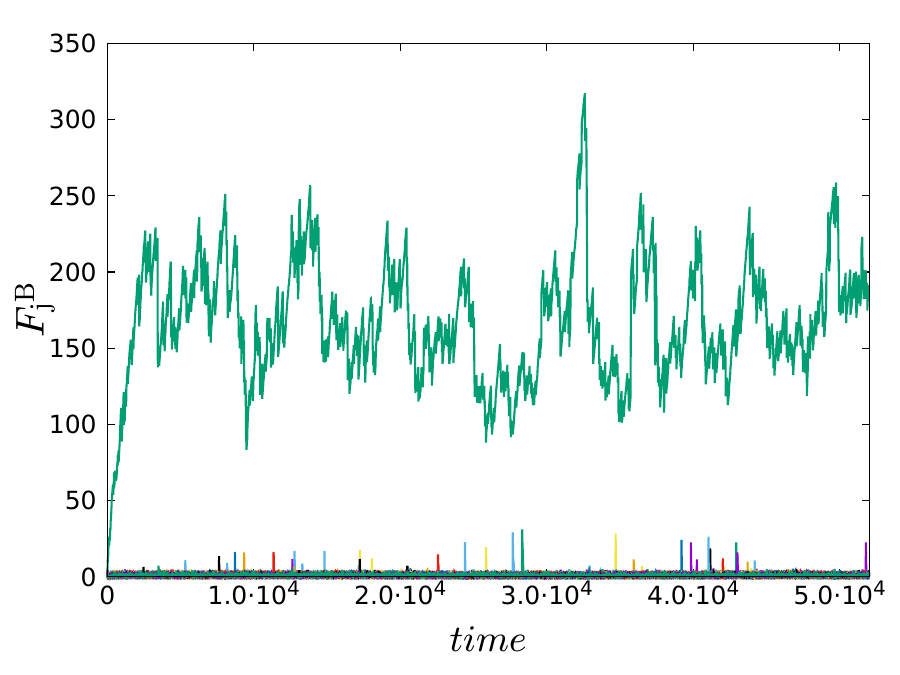}
              \caption{}
        \end{subfigure}
        \begin{subfigure}[b]{0.45\textwidth}
		\includegraphics[width=\textwidth]{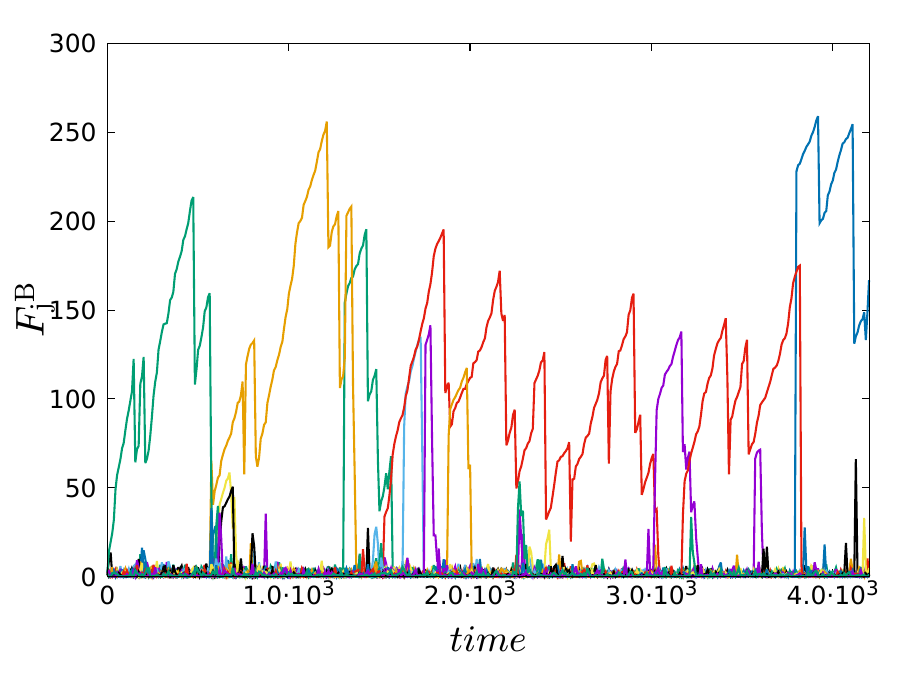}
              \caption{}
        \end{subfigure}
		\caption{\textbf{Temporal evolution of the fitnesses of group B with two values of $x$.} Time evolution of all $F_{j}^\text{B}$ for $N_\text{A}=500$ and $N_\text{B}=50$ individuals, $\eta= 5$, randomly simulated on a $L\times L$ lattice ($L=25$) and two different exchange factors. (a) $x=0.10$. (b) $x=0.50$.}
	\label{f:x-L25-o}
\end{figure}

Below, we present the relationship between the temporal evolution of fitness and the exchange factor denoted as $x$. Figure \ref{f:x-L25-o} illustrates the temporal evolution of fitnesses of group B during the stationary regime for two distinct values of $x$, while maintaining the same parameters as Figure \ref{f:5-001-L25}. As discussed in the main text, a comparison between Figure \ref{f:5-001-L25} (b) and Figure \ref{f:x-L25-o} (a) reveals that the time taken to reach the stationary regime significantly decreases with increasing $x$, albeit with a notable increase in stochasticity within this regime. With further increments in $x$, as depicted in Figure \ref{f:x-L25-o} (b), the exchange factor becomes sufficiently large to result in the emergence of agents with lower fitness levels upon encounters with more powerful agents of the opposing group A, and vice versa.

%\newpage
\bibliographystyle{model1-num-names}
%\bibliographystyle{cas-model2-names}
\bibliography{cas-refs}
%\bibliographystyle{unsrtnat}


%\newpage

\end{document}